\documentclass[11pt]{article}
\usepackage{epsfig} 
\newcommand{\sect}[1]{ \section{#1} \setcounter{equation}{0} }

\newcommand{\pslash}{p \! \! \! /} 
\newcommand{\qslash}{q \! \! \! /}

\newcommand{\third}{\mbox{\small{$\frac{1}{3}$}}} 
 
\newcommand{\pitwo}{\mbox{\small{$\frac{\pi}{2}$}}} 
\newcommand{\pisix}{\mbox{\small{$\frac{\pi}{6}$}}} 
\newcommand{\MSbar}{\overline{\mbox{MS}}} 
\newcommand{\MSbars}{\overline{\mbox{\footnotesize{MS}}}} 
\newcommand{\MOMgggs}{\mbox{\footnotesize{MOMggg}}}
\newcommand{\MOMhs}{\mbox{\footnotesize{MOMh}}}
\newcommand{\MOMqs}{\mbox{\footnotesize{MOMq}}}
\newcommand{\MOMis}{\mbox{\footnotesize{MOMi}}}
\newcommand{\RIps}{\mbox{\footnotesize{RI${}^\prime$}}}
\newcommand{\Nf}{N_{\!f}}

\begin{document}
\title{Two loop QCD vertices at the symmetric point}
\author{J.A. Gracey, \\ Theoretical Physics Division, \\ 
Department of Mathematical Sciences, \\ University of Liverpool, \\ P.O. Box 
147, \\ Liverpool, \\ L69 3BX, \\ United Kingdom.} 
\date{} 
\maketitle 

\vspace{5cm} 
\noindent 
{\bf Abstract.} We compute the triple gluon, quark-gluon and ghost-gluon
vertices of QCD at the symmetric subtraction point at two loops in the $\MSbar$
scheme. In addition we renormalize each of the three vertices in their 
respective momentum subtraction schemes, MOMggg, MOMq and MOMh. The conversion 
functions of all the wave functions, coupling constant and gauge parameter 
renormalization constants of each of the schemes relative to $\MSbar$ are 
determined analytically. These are then used to derive the {\em three} loop 
anomalous dimensions of the gluon, quark, Faddeev-Popov ghost and gauge 
parameter as well as the $\beta$-function in an arbitrary linear covariant 
gauge for each MOM scheme. There is good agreement of the latter with earlier 
Landau gauge numerical estimates of Chetyrkin and Seidensticker.  

\vspace{-17.2cm}
\hspace{13.5cm}
{\bf LTH 921}

\newpage

\sect{Introduction.} 

The structure of the renormalization group functions of Quantum Chromodynamics 
(QCD) has been established to a high degree of precision for well over a decade
now. Originally the one loop discovery of asymptotic freedom due to the 
negative $\beta$-function, \cite{1,2}, was swiftly followed by the two loop 
computation, \cite{3,4}. Within a decade the three loop term emerged, \cite{5},
but the four loop result, \cite{6}, took substantially longer to determine. It 
was later confirmed in \cite{7}. To appreciate how involved the calculation of
\cite{6} was, it required the evaluation of the order of $50000$ of Feynman 
diagrams and an intense amount of symbolic algebraic manipulation using the 
language {\sc Form}, \cite{8}. Also new techniques to evaluate four loop 
Feynman integrals were developed, \cite{8}, as well as the coding of a 
{\sc Form} routine to handle the colour group algebra manipulations 
automatically, \cite{9}. In addition to the $\beta$-function governing the 
running coupling constant the other renormalization group functions have been 
available at a variety of loop orders over the same time scale, 
\cite{10,11,12,13,14,15}. In summarizing the status of the basic 
renormalization group functions of QCD we have concentrated on the status of
the modified minimal subtraction, $\MSbar$, scheme. This renormalization scheme
is the most commonly used for perturbative QCD computations. It is used
primarily because it is a mass independent scheme. Consequently one simplifying
feature is that all the Feynman graphs which one needs to evaluate to determine
the $\beta$-function to four loops in this scheme involve single scale 
$2$-point Feynman integrals. This means, for instance, that techniques such as 
integration by parts can be systematically used to reduce integrals to either 
basic master topologies or simple chain type integrals whose evaluation is 
effectively trivial. Indeed the {\sc Mincer} algorithm, \cite{16}, is an 
excellent example of the implementation of integration by parts and has been 
coded in {\sc Form}, \cite{8}. Briefly {\sc Mincer} determines massless 
$2$-point integrals at three loops to the finite part with respect to the 
regulator. The divergences are written as simple poles in $\epsilon$ where 
$d$~$=$~$4$~$-$~$2\epsilon$ and $d$ is the arbitrary spacetime dimension of 
dimensional regularization. The latter, which we will use, is the main 
regularization method for perturbative quantum field theories and, for 
instance, preserves gauge invariance. Whilst {\sc Mincer} has been used
extensively for many problems other than the basic renormalization group
functions of QCD, the {\sc Form} version, \cite{17}, was used to renormalize 
the QCD coupling constant at three loops, \cite{12}. The reason for this is 
that one can nullify an external momentum of one of the legs of a $3$-point 
function, without introducing spurious infrared infinities, and hence apply the
{\sc Mincer} algorithm to obtain the correct three loop $\MSbar$ 
$\beta$-function of \cite{5}. Although this has proved to be a powerful 
technique for renormalization group functions, for more physical problems the 
$\MSbar$ scheme has several drawbacks.  
 
One of these is the fact that the $\MSbar$ scheme is not a physical scheme.
Specifically, the subtraction at the vertex as noted earlier in the 
{\sc Mincer} approach is at a point of exceptional momentum. Whilst this is
sufficient to extract the divergences and hence find the renormalization 
constant, one would have potential infrared ambiguities if one were trying to 
make a measurement of the finite part of the associated vertex function at this 
exceptional momentum configuration. This has been known for a long time,
\cite{18,19}, but has also been remarked upon again more recently in the 
specific context of Green's functions with an inserted operator, \cite{20}. 
Indeed the measurement of vertices is a topic of interest as one can study any 
of the vertices non-perturbatively using say lattice regularization. Such 
analyses are necessary for determining the strength of the strong coupling 
constant accurately for comparison with experiment and examining its behaviour 
at low energy. Therefore, it seems appropriate to consider vertex momentum
configurations which are non-exceptional and hence schemes which are physical
rather than the unphysical minimal schemes. A set of such schemes was
introduced in \cite{21} and given the general designation of momentum 
subtraction schemes being denoted by MOM. In essence they consist of two
criteria. The first is that $3$-point vertices are considered at a symmetric
subtraction point, \cite{21}. In other words the squares of the three external 
momenta are all set equal to each other. Therefore, there is no nullification 
of an external momentum and hence no exceptional momenta. The second is that at
the subtraction point the scheme is defined in such a way that after 
renormalization there are no $O(a)$ corrections where $a$~$=$~$g^2/(16\pi^2)$
and $g$ is the gauge coupling constant. Thus the renormalization constants all 
contain finite parts, \cite{21}, in addition to the poles in $\epsilon$ which 
must always be subtracted. In \cite{21} the MOM schemes were analysed 
comprehensively at one loop. By schemes we mean the three types derived from 
the respective three $3$-point vertices in the canonical linear covariant gauge
fixing. These are denoted by MOMggg, MOMh and MOMq due to their origin from the 
respective triple gluon, ghost-gluon and quark-gluon vertices. Indeed these one
loop MOM scheme computations of the renormalization group functions have been 
the state of the art for a long time for a relatively simple reason. This is 
because for the vertex renormalization the symmetric subtraction point momentum 
configuration introduces single scale $3$-point Feynman integrals, \cite{21}. 
At one loop there is only one such basic master integral to evaluate which was 
given in \cite{21}. However, at two loops there are several basic master 
integrals which were only evaluated analytically in recent years. Therefore, it
is the purpose of this article to extend the work of \cite{21} to two loops. We
will achieve this by evaluating all $2$-point and $3$-point vertices to the 
finite part at two loops at the symmetric subtraction point analytically. Hence
we will determine the gluon, Faddeev-Popov ghost and quark wave function and 
coupling constant renormalization constants for each of the three MOM schemes 
in an arbitrary linear covariant gauge. En route we will also provide the 
structure of each of the three vertices to the finite part in the $\MSbar$ 
scheme at the symmetric point. This information should prove useful to lattice
groups seeking to measure any of the vertex functions in the $\MSbar$ scheme.
For instance, in such an exercise the lattice results must match onto the 
ultraviolet part of the vertex function which is where perturbation theory is
valid. Equipped with the two loop renormalization constants we will also 
determine the conversion functions for each scheme including the relation 
between the coupling constants. Hence via properties of the renormalization 
group we will determine each MOM scheme renormalization group function to 
{\em three} loops in an arbitrary linear covariant gauge. 

Finally, we mention other related work in this area. Prior to this article 
there were approximate calculations of the three MOM $\beta$-functions in the 
Landau gauge, \cite{22}. This was achieved by approximating the basic master 
two loop integrals at the symmetric point by an expansion where one of the 
external momentum is marked to produce an asymmetric Green's function. Then the
integrals were expanded in the ratio of this marked momentum to the other
independent momentum, \cite{22,23}. So if a sufficient number of terms is 
computed in this parameter, where the coefficients are rapidly decreasing in
size, then a reasonable approximation can be deduced by truncating at an 
appropriate order, \cite{22,23}. Moreover, error estimates can be deduced. In 
this expansion the $3$-point functions reduce to $2$-point functions. Hence 
each term in the series is evaluated using the {\sc Mincer} algorithm. Indeed 
we previewed the results of this article in \cite{24}, by providing the exact 
three loop QCD MOM $\beta$-functions in the Landau gauge and demonstrated how 
accurate the results were in comparison to \cite{22}. This was a very close 
overlap which was impressive given the level of computing technology available 
to the authors of \cite{22} at that time. Next, it would be remiss if we failed
to mention a related three loop MOM $\beta$-function computation given in 
\cite{25}. There the MOM $\beta$-function was defined using the invariant 
charge concept of \cite{26,27}. As it involves only finite parts of $2$-point 
functions it is independent of any of the vertices of the theory unlike the 
MOMggg, MOMh or MOMq schemes. Hence, aside from the Riemann zeta series, it 
does not have any of the special number structures, such as harmonic 
polylogarithms, which derive from the single scale $3$-point master integrals 
and are evident in our analytic results, \cite{24}. Also in this context a 
variation on this theme of establishing MOM scheme definitions based solely on 
$2$-point functions has been developed more recently in \cite{28}. Known as the
minimal MOM scheme it circumvents a renormalization condition based on the 
ghost-gluon vertex $3$-point function by imposing the alternative condition 
that the renormalization constant of the ghost-gluon vertex is the {\em same} 
as that of the vertex in the $\MSbar$ scheme itself. One benefit of this is 
that in a linear covariant gauge a non-perturbative running coupling constant 
can be defined purely in terms of the gluon and ghost propagator form factors. 
With this definition the running coupling constant can be measured more 
accurately in principle using lattice gauge theory techniques. For instance, 
recent activity on this specific aspect can be found in \cite{29,30}. Given 
that one can determine the gluon and ghost $2$-point functions to several loop 
orders in perturbation theory, the four loop minimal MOM $\beta$-function has 
been determined in \cite{28}. Though unlike \cite{25} quark mass effects have 
only been estimated in \cite{28}. Finally, we note that the measurement of 
vertices non-perturbatively is not exclusively studied by lattice techniques. 
For instance, recently the triple gluon vertex was examined using 
Schwinger-Dyson methods, \cite{31}. Therefore, the structure of the two loop 
vertices given at the symmetric point here should also prove relevant in 
analyses using those techniques too. Other two loop studies of $3$-point 
vertices of QCD include the results of \cite{32,33} where the triple gluon 
vertex was examined in the zero momentum limit and in the on-shell 
configuration respectively. The latter also includes the same analysis for the 
ghost-gluon vertex. 

The article is organized as follows. We outline the general formalism and
notation we use in section $2$ as well as discussing various aspects of the 
renormalization in each of the three MOM schemes in section $3$. The results 
for the $\MSbar$ and MOM amplitudes of the respective vertex functions as well
as the conversion functions and the three loop renormalization group functions 
are given in each of the three following sections\footnote{All the results
presented in the article, including the full analytic forms for an arbitrary
gauge, have been included in an attached electronic data file for each of the 
three MOM schemes.}. Finally, we present conclusions in section $7$. An 
appendix contains the explicit tensors of the bases for each of the three 
vertices as well as the respective projection matrices.

\sect{Preliminaries.}

We begin by discussing the general features of the computation we perform. The
three Green's functions we consider are $\left\langle A^a_\mu(p) A^b_\nu(q) 
A^c_\sigma(r) \right\rangle$, $\left\langle \psi^i(p) \bar{\psi}^j(q) 
A^c_\sigma(r) \right\rangle$ and $\left\langle c^a(p) \bar{c}^b(q) 
A^c_\sigma(r) \right\rangle$ where $r$~$=$~$-$~$p$~$-$~$q$ by momentum
conservation. The symmetric subtraction point is defined by the condition
\begin{equation}
p^2 ~=~ q^2 ~=~ r^2 ~=~ -~ \mu^2
\label{symmpt}
\end{equation}
where $\mu$ is the common mass scale. It will also be used as the mass scale to
ensure that the coupling constant remains dimensionless in dimensional 
regularization in $d$-dimensions which we use throughout. Therefore our results
for the finite parts of the vertex functions will not involve logarithms which
can be restored from knowledge of the renormalization group functions. 
From (\ref{symmpt}) we have
\begin{equation}
pq ~=~ \frac{1}{2} \mu^2
\end{equation}
and we will use $p$ and $q$ as the two independent momenta throughout. Their 
sum will be taken to flow out through a gluon external leg which will
therefore be the reference leg in each of the vertices. Given that each vertex 
has a colour group tensor associated with it, we factor it off when we consider
the symmetric point, which we do exclusively from now on, by defining 
\begin{eqnarray}
\left. \left\langle A^a_\mu(p) A^b_\nu(q) A^c_\sigma(-p-q)
\right\rangle \right|_{p^2 = q^2 = - \mu^2} &=& f^{abc} 
\left. \Sigma^{\mbox{\footnotesize{ggg}}}_{\mu \nu \sigma}(p,q)
\right|_{p^2 = q^2 = - \mu^2} \nonumber \\
\left. \left\langle \psi^i(p) \bar{\psi}^j(q) A^c_\sigma(-p-q)
\right\rangle \right|_{p^2 = q^2 = - \mu^2} &=& T^c_{ij} 
\left. \Sigma^{\mbox{\footnotesize{qqg}}}_\sigma(p,q)
\right|_{p^2 = q^2 = - \mu^2} \nonumber \\
\left. \left\langle c^a(p) \bar{c}^b(q) A^c_\sigma(-p-q)
\right\rangle \right|_{p^2 = q^2 = - \mu^2} &=& f^{abc} 
\left. \Sigma^{\mbox{\footnotesize{ccg}}}_\sigma(p,q)
\right|_{p^2 = q^2 = - \mu^2} ~. 
\label{vertdef}
\end{eqnarray}
We will use ggg, qqg and ccg in equations to denote the triple gluon,
quark-gluon and ghost-gluon vertex functions respectively. Next we decompose
the Lorentz amplitudes 
$\left. \Sigma^{\mbox{\footnotesize{ggg}}}_{\mu \nu \sigma}(p,q)
\right|_{p^2 = q^2 = - \mu^2}$,  
$\left. \Sigma^{\mbox{\footnotesize{qqg}}}_\sigma(p,q)
\right|_{p^2 = q^2 = - \mu^2}$ and 
$\left. \Sigma^{\mbox{\footnotesize{ccg}}}_\sigma(p,q)
\right|_{p^2 = q^2 = - \mu^2}$ into the scalar amplitudes. The Lorentz tensor 
basis for each function is not the same. Nor is the choice of basis we will use
for the decomposition unique. The explicit forms of the tensors are given in 
Appendix A. Though we note that away from the symmetric point, where the 
equalities of (\ref{symmpt}) are no longer valid, then the basis will involve a
larger number of tensors. Therefore, we formally define the scalar amplitudes 
as  
\begin{eqnarray}
\left. \frac{}{} \Sigma^{\mbox{\footnotesize{ggg}}}_{\mu \nu \sigma}(p,q)
\right|_{p^2 = q^2 = - \mu^2} &=& \sum_{k=1}^{14}
{\cal P}^{\mbox{\footnotesize{ggg}}}_{(k) \, \mu \nu \sigma }(p,q) \,
\Sigma^{\mbox{\footnotesize{ggg}}}_{(k)}(p,q) \nonumber \\
\left. \frac{}{} \Sigma^{\mbox{\footnotesize{qqg}}}_\sigma(p,q)
\right|_{p^2 = q^2 = - \mu^2} &=& \sum_{k=1}^{6}
{\cal P}^{\mbox{\footnotesize{qqg}}}_{(k) \, \sigma }(p,q) \,
\Sigma^{\mbox{\footnotesize{qqg}}}_{(k)}(p,q) \nonumber \\
\left. \frac{}{} \Sigma^{\mbox{\footnotesize{ccg}}}_\sigma(p,q)
\right|_{p^2 = q^2 = - \mu^2} &=& \sum_{k=1}^{2}
{\cal P}^{\mbox{\footnotesize{ccg}}}_{(k) \, \sigma }(p,q) \,
\Sigma^{\mbox{\footnotesize{ccg}}}_{(k)}(p,q)
\label{ampdecmp}
\end{eqnarray}
where $\Sigma^i_{(k)}(p,q)$ are the scalar amplitudes at (\ref{symmpt}) and 
${\cal P}^{i ~\, \mu_1 \ldots \mu_{n_i}}_{(k)}(p,q)$ are the tensors of the 
respective bases with $i$ corresponding to one of ggg, qqg or ccg. We use $k$
to label the basis elements and have chosen the labelling in such a way that 
channel $1$ corresponds to the tensor which occurs in the Feynman rule of the 
corresponding vertex in the QCD Lagrangian. Though given the structure of the 
triple gluon vertex the first six tensors are part of that vertex.

To determine the values of each of the scalar amplitudes we use the method of 
projection. In other words an identified amplitude can be isolated by 
multiplying the Green's function by a specific linear combination of the basis 
tensors. Therefore, we have 
\begin{eqnarray}
f^{abc} \Sigma^{\mbox{\footnotesize{ggg}}}_{(k)}(p,q) &=& 
{\cal M}^{\mbox{\footnotesize{ggg}}}_{kl} \left(
{\cal P}^{\mbox{\footnotesize{ggg}} \, \mu \nu \sigma}_{(l)}(p,q) \left. 
\left\langle A^a_\mu(p) A^b_\nu(q) A^c_\sigma(-p-q) 
\right\rangle \right )\right|_{p^2 = q^2 = - \mu^2} 
\nonumber \\
T^c_{ij} \Sigma^{\mbox{\footnotesize{qqg}}}_{(k)}(p,q) &=& 
{\cal M}^{\mbox{\footnotesize{qqg}}}_{kl} \left(
{\cal P}^{\mbox{\footnotesize{qqg}} \, \sigma}_{(l)}(p,q) \left. 
\left\langle \psi^i(p) \bar{\psi}^j(q) A^c_\sigma(-p-q) 
\right\rangle \right) \right|_{p^2 = q^2 = - \mu^2} 
\nonumber \\
f^{abc} \Sigma^{\mbox{\footnotesize{ccg}}}_{(k)}(p,q) &=& 
{\cal M}^{\mbox{\footnotesize{ccg}}}_{kl} \left(
{\cal P}^{\mbox{\footnotesize{ccg}} \, \sigma}_{(l)}(p,q) \left. 
\left\langle c^a(p) \bar{c}^b(q) A^c_\sigma(-p-q) 
\right\rangle \right) \right|_{p^2 = q^2 = - \mu^2} 
\end{eqnarray}
for each vertex where we have included the passive colour factor on the left
hand side to complement the one which is implicit on the right side. The free
spinor indices in the quark-gluon vertex have been left implicit. The matrix
${\cal M}^i_{kl}$ is the projection matrix and the explicit forms for each of 
the vertices are given in the appendix. It is computed by first finding the 
matrix ${\cal N}^i_{kl}$ for each vertex which is constructed from the basis 
tensors by Lorentz contraction in $d$-dimensions using the conditions of the 
symmetric point, (\ref{symmpt}). In other words  
\begin{equation}
{\cal N}^i_{kl} ~=~ \left. \left( {\cal P}^i_{(k) \, \mu_1 \ldots 
\mu_{n_i}}(p,q) {\cal P}^{i ~\, \mu_1 \ldots \mu_{n_i}}_{(l)}(p,q) \right)
\right|_{p^2=q^2=-\mu^2}.
\end{equation}
This produces a matrix, ${\cal N}^i_{kl}$, whose entries are polynomials in $d$
and ${\cal M}^i_{kl}$ corresponds to its inverse. For the quark-gluon vertex 
the tensor basis necessarily has to be built from $\gamma$-matrices in addition
to the momentum vectors. As we will be working in dimensional regularization we
will use the generalized $\gamma$-matrices,
$\Gamma_{(n)}^{\mu_1 \ldots \mu_n}$, \cite{34,35,36}. These form a complete set
of matrices which span the spinor space of the associated $d$-dimensional 
spacetime. They are defined to be completely antisymmetric in the Lorentz 
indices and are given by  
\begin{equation}
\Gamma_{(n)}^{\mu_1 \ldots \mu_n} ~=~ \gamma^{[\mu_1} \ldots \gamma^{\mu_n]}
\end{equation}
where an overall factor of $1/n!$ is understood and $\Gamma_{(0)}$ is the unit
element. One beneficial property, among other general properties \cite{37,38},
is that the trace over the generalized $\gamma$-matrices is isotropic as
\begin{equation}
\mbox{tr} \left( \Gamma_{(m)}^{\mu_1 \ldots \mu_m}
\Gamma_{(n)}^{\nu_1 \ldots \nu_n} \right) ~ \propto ~ \delta_{mn}
I^{\mu_1 \ldots \mu_m \nu_1 \ldots \nu_n}
\end{equation}
and $I^{\mu_1 \ldots \mu_m \nu_1 \ldots \nu_n}$ is the unit tensor. For the 
quark-gluon vertex only the $n$~$=$~$1$ and $3$ generalized matrices arise, as 
we are working in the chiral limit throughout, which can be seen in the 
explicit decomposition in appendix A. 

We have used several main working tools to complete our analysis which are all
computer based as it would be virtually impossible to proceed without automatic
Feynman diagram generators as well as symbolic manipulation programmes. For
each of the three vertex functions the Feynman graphs are constructed with the
{\sc Qgraf} package, \cite{39}. For the triple gluon vertex there are $8$ one
loop and $106$ two loop diagrams contributing to the Green's function. For the
other two vertices the number of graphs is the same with $2$ one loop and $33$ 
two loop diagrams. From the {\sc Qgraf} output the Lorentz and colour indices 
are appended in the symbolic manipulation language {\sc Form}, \cite{8}. Indeed
{\sc Form} is used as the machinery for the rest of our algebraic computations 
as it is efficient in handling the huge amounts of algebra required for our 
analysis. To compute the Feynman graphs to the required finite part in 
dimensional regularization we use the Laporta algorithm, \cite{40}. Briefly the 
aim is to write each Feynman diagram in the Green's function in terms of a set
of basic scalar master integrals whose expressions to the finite part are
known. After applying the projection matrix to the Green's function the 
resulting scalar Feynman integrals are written in a specific format. Given the 
structure of the QCD propagators and vertices the scalar products in the 
numerators are rewritten in terms of the propagator denominators in preparation
for using the method of \cite{40}. However, given the symmetric point condition
then for all the topologies there will be irreducible numerators where there 
are no corresponding denominators. Equally there will be propagators raised to 
a power larger than unity. To reduce this very large number of scalar integrals
to the set of masters requires an intense amount of integration by parts. One 
method which achieves this is the Laporta approach, \cite{40}. This 
systematically determines all the integration by parts relations, as well as 
Lorentz identity relations, between all the integrals which are needed. The
algorithm then uses a systematic way of reducing integrals classified in 
various levels to the lower levels or to a basic master in that level. The 
beauty of the method is that it terminates and can be coded for implementation 
on a computer. There are several available packages. We have chosen to use 
{\sc Reduze}, \cite{41}, which is written in the symbolic manipulation 
formalism of {\sc GiNaC}, \cite{42}, whose working language is C++. One main 
aspect of the package which we exploit is to construct a database of relations 
between the integrals and then to lift out those we require for our specific 
computation. These are simply mapped to {\sc Form} format and an integration 
module included within our overall automatic {\sc Form} programme. This sums up
all the contributions to each of the Green's functions allowing us to perform 
the renormalization in any of the schemes of interest. For the latter we follow
the algorithm for automatic Feynman diagram computations devised in \cite{12}. 
This involves performing all the integrals as a function of bare parameters 
with the renormalized values introduced by a simple rescaling via the 
respective renormalization constants. Therefore, this means that we only 
compile the results for the vertex amplitudes once prior to determining the 
$\MSbar$ or MOM scheme renormalization group functions and amplitudes. 

Finally, as the analytic results we derive involve cumbersome expressions even
in the Landau gauge as will be evident, we will give numerical values for all
our results in an arbitrary gauge. As our computations revolve around the 
symmetric subtraction point the underlying one and two loop scalar master 
integrals involve structures not seen in the $3$-point momentum configuration 
where one external momentum is nullified. In the latter case at low loop order 
one ordinarily only encounters rationals and the Riemann zeta function, 
$\zeta(z)$. For the symmetric point one finds the function 
\begin{equation}
s_n(z) ~=~ \frac{1}{\sqrt{3}} \Im \left[ \mbox{Li}_n \left(
\frac{e^{iz}}{\sqrt{3}} \right) \right] 
\end{equation}
for various arguments where $\mbox{Li}_n(z)$ is the polylogarithm function as
well as one specific combination of the harmonic polylogarithms which we denote
by
\begin{equation}
\Sigma ~=~ {\cal H}^{(2)}_{31} ~+~ {\cal H}^{(2)}_{43} ~.
\end{equation}
The master integrals where these originally arise are summarized in \cite{43}
but the explicit evaluation are given in a set of articles, \cite{44,45,46,47}.
Given that these functions will arise we record the relevant numerical values
that we needed which are
\begin{eqnarray}
\zeta(3) &=& 1.20205690 ~~,~~ \Sigma ~=~ 6.34517334 ~~,~~
\psi^\prime\left( \frac{1}{3} \right) ~=~ 10.09559713 ~, \nonumber \\
\psi^{\prime\prime\prime}\left( \frac{1}{3} \right) &=& 488.1838167 ~~,~~
s_2\left( \frac{\pi}{2} \right) ~=~ 0.32225882 ~~,~~
s_2\left( \frac{\pi}{6} \right) ~=~ 0.22459602 ~, \nonumber \\
s_3\left( \frac{\pi}{2} \right) &=& 0.32948320 ~~,~~
s_3\left( \frac{\pi}{6} \right) ~=~ 0.19259341 
\end{eqnarray}
where $\psi(z)$ is the derivative of the logarithm of Euler $\Gamma$-function. 

\sect{Renormalization.}

We devote this section to general aspects of MOM scheme renormalization. Having
described the computer algebraic machinery used to construct the vertex 
functions we now recall how the MOM schemes are defined where we regard the
amplitudes of (\ref{ampdecmp}) as having been determined to the finite part. 
The divergences are removed into the coupling constant renormalization 
constant. However, to two loops this is an iterative procedure which is 
entwined with the $2$-point function renormalization. This is because to 
extract the coupling constant counterterm from the vertex function one has to 
pay attention to the wave function renormalization constants of the external 
fields of the vertex function. Therefore, one first determines the one loop 
wave function renormalization constants in the MOM scheme of interest which is 
then fixed in examining one loop vertex function defining that particular MOM 
scheme. For both the $2$-point functions and the specific vertex the MOMi 
renormalization constant is defined so that at the subtraction point there are
no $O(a)$ corrections after the renormalization constant is defined. Throughout
we will use the syntax that in MOMi or equations $\mbox{i}$ represents ggg, q 
or h.  Once the one loop renormalization constants are fully determined then 
one repeats the exercise for the two loop contribution to first the $2$-point 
function and then the associated vertex. The reason for explicitly defining the
procedure is to ensure that there is no inconsistency in determining the 
coupling constant renormalization constant. The finite parts of the one loop 
wave function renormalization constants impact upon the finite parts of the two
loop MOMi coupling constant renormalization constants, \cite{21}. Otherwise an 
inconsistency in the renormalization group would emerge in trying to deduce the
anomalous dimensions as well as the associated conversion functions for each 
renormalization constant. As a check on the vertex functions we have computed,
we have verified that the two loop $\MSbar$ coupling constant renormalization 
constant of \cite{1,2,3,4} correctly emerges when renormalizing at the 
symmetric subtraction point. One final point concerning our MOMi scheme 
renormalizations and that is that we first determine all the amplitudes before 
setting the renormalization constants for each scheme. For MOMi schemes we 
render that channel with the $\epsilon$ divergences to have no $O(a)$ 
corrections. However, we emphasise that this is by no means the only way of 
defining the renormalization constants within the MOM ethos. An alternative, 
for instance, is to first multiply the vertex function with its Lorentz tensor 
structure present, by the tensor of the corresponding Feynman rules. Then the 
coupling constant renormalization constant is defined by ensuring that there 
are no $O(a)$ corrections to this object. We note that doing this for each of 
the three schemes would involve several of the non-Feynman rule amplitudes in 
contributing to the coupling constant renormalization. We have not proceeded in
this way as it does not appear to be in keeping with \cite{21,22}. However, in 
providing the full vertex structure in terms of the Lorentz tensors in the 
$\MSbar$ scheme an interested reader has the opportunity to study such 
alternative MOM scheme definitions. Indeed it may be the case that convergence 
of certain perturbative series could be improved in such a way. 

Having reviewed the procedure we followed we now comment on the relation of the
parameters of the theory in different schemes. For QCD the relevant parameters
are the coupling constant and the linear covariant gauge fixing parameter. As
the former is vertex dependent we comment on the latter first. As outlined the
MOMi renormalizations require $2$-point function renormalization. Therefore,
like the coupling constant, \cite{21}, the gauge parameter can be different in 
different schemes. To relate them we follow the standard method and define
\begin{equation}
\alpha_{\MOMis}(\mu) ~=~ \frac{Z_A^{\MOMis}}{Z_A^{\MSbars}}
\alpha_{\MSbars}(\mu) 
\label{aldef}
\end{equation}
where the subscript on the parameter refers to the scheme the variable is
defined with respect to and $Z_A$ is the gluon wave function renormalization
constant. We follow the same conventions as \cite{48} in defining the gauge
parameter renormalization constant, $Z_\alpha$, by 
\begin{equation}
\alpha_{\mbox{\footnotesize{o}}} ~=~ Z^{-1}_\alpha Z_A \, \alpha
\end{equation}
where the subscript, ${}_{\mbox{\footnotesize{o}}}$, indicates the bare 
parameter. With this convention then the respective anomalous dimensions 
satisfy, \cite{48},  
\begin{equation}
\gamma_\alpha(a,\alpha) ~=~ -~ \gamma_A(a,\alpha) 
\end{equation}
in each scheme and $\gamma_\alpha(a,\alpha)$ is the anomalous dimension of the 
linear covariant gauge parameter. In carrying out the renormalization in each 
of the three schemes we have determined $\alpha_{\MOMis}(\mu)$ for each of the 
three cases and found that using the MOM scheme definition of \cite{21}  
\begin{eqnarray}
\alpha_{\MOMis} &=& \left[ 1 + \left[ \left[ 80 T_F \Nf - 9 \alpha_{\MSbars}^2 
- 18 \alpha_{\MSbars} - 97 \right] C_A \right] \frac{a_{\MSbars}}{36} \right. 
\nonumber \\
&& \left. \,+ \left[ \left[ 18 \alpha_{\MSbars}^4 - 18 \alpha_{\MSbars}^3
+ 190 \alpha_{\MSbars}^2 - 576 \zeta(3) \alpha_{\MSbars} + 463 \alpha_{\MSbars}
+ 864 \zeta(3) - 7143 \right] C_A^2 \right. \right. \nonumber \\
&& \left. \left. ~~~~~- \left[ 320 \alpha_{\MSbars}^2 + 320 \alpha_{\MSbars} 
- 2304 \zeta(3) - 4248 \right] C_A T_F \Nf \right. \right. \nonumber \\
&& \left. \left. ~~~~~- \left[ 4608 \zeta(3) - 5280 \right] C_F T_F \Nf \right]
\frac{a^2_{\MSbars}}{288} ~+~ O \left( a^3_{\MSbars} \right) \right] 
\alpha_{\MSbars} ~. 
\label{almap}
\end{eqnarray}
To ensure that the mapping is not divergent due to poles in $\epsilon$ one has 
to iteratively solve (\ref{aldef}) order by order in perturbation theory hand 
in hand with the coupling constant of that scheme. This is because when we set 
the renormalization constants in a scheme both parameters of the explicit forms
belong to that particular scheme. From (\ref{almap}) we see that the gauge 
parameter mapping is the same for all three schemes. Indeed it is the same as 
that for the RI${}^\prime$ scheme, \cite{48}. This is not unexpected as the 
parameter mapping is effectively the conversion function and reflects an 
underlying feature of the renormalization group. In essence it tracks how the 
scheme is defined for that one parameter amidst the renormalization of all the 
other parameters and wave functions within the Green's functions. This will 
become evident later for other renormalizations. Therefore, given this feature 
the three loop term of (\ref{almap}) has already been given in \cite{48}.
Further, we recall that it means that the Landau gauge is preserved between the
schemes.  

The procedure to define the relation between the coupling constants in 
different schemes is similar. Though as we are dealing with three MOM schemes
then the definitions are different in each case. We follow \cite{21} and 
\cite{22} for this and define
\begin{eqnarray}
a_{\MOMgggs}(\mu) &=& a_{\MSbars}(\mu) \left. \left[ \frac{\Pi_g^{\MOMgggs}(p)}
{\Pi_g^{\MSbars}(p)} \right]^3 \right|_{p^2 \, = \,-\, \mu^2} \!\!
\left[ \frac{\Sigma^{\mbox{\footnotesize{ggg}}}_{(1)\,\MSbars}(-\mu^2,-\mu^2)}
{\Sigma^{\mbox{\footnotesize{ggg}}}_{(1)\,\MOMgggs}(-\mu^2,-\mu^2)} \right]^2 
\nonumber \\
a_{\MOMqs}(\mu) &=& a_{\MSbars}(\mu) \left. \left[ \frac{\Pi_g^{\MOMqs}(p) 
\left( \Sigma^{\MOMqs}_q(p) \right)^2} {\Pi_g^{\MSbars}(p) \left( 
\Sigma^{\MSbars}_q(p) \right)^2} \right] \right|_{p^2 \, = \,-\, \mu^2} \!\!
\left[ \frac{\Sigma^{\mbox{\footnotesize{qqg}}}_{(1)\,\MSbars}(-\mu^2,-\mu^2)}
{\Sigma^{\mbox{\footnotesize{qqg}}}_{(1)\,\MOMqs}(-\mu^2,-\mu^2)} \right]^2 
\nonumber \\
a_{\MOMhs}(\mu) &=& a_{\MSbars}(\mu) \left. \left[ \frac{\Pi_g^{\MOMhs}(p) 
\left( \Sigma^{\MOMhs}_c(p) \right)^2} {\Pi_g^{\MSbars}(p) \left( 
\Sigma^{\MSbars}_c(p) \right)^2} \right] \right|_{p^2 \, = \,-\, \mu^2} \!\!
\left[ \frac{\Sigma^{\mbox{\footnotesize{gcc}}}_{(1)\,\MSbars}(-\mu^2,-\mu^2)}
{\Sigma^{\mbox{\footnotesize{gcc}}}_{(1)\,\MOMhs}(-\mu^2,-\mu^2)} \right]^2 
\label{ccdefs}
\end{eqnarray}
where we use the same designation for the vertices as before. Clearly the
definitions involve the respective vertex functions evaluated at the symmetric
subtraction point. Moreover, the amplitude chosen is that which has the
divergences in $\epsilon$ prior to renormalization or equivalently the 
amplitude which corresponds to the vertex Feynman rule. In the case of the
triple gluon vertex we have chosen channel $1$ which is only part of the 
Feynman rule. However, as will be apparent from the explicit results the other 
amplitudes from $2$ to $6$ are related in the way one would expect from the
vertex structure so that our definition is consistent. Whilst in each of the
three definitions the corresponding amplitude in the MOMi schemes have no 
$O(a)$ corrections, we have formally included it to ensure the normalization is
correct and that the ratio of the vertex amplitudes from MOMi to $\MSbar$ 
begins with unity. The other main feature of (\ref{ccdefs}) is the presence of 
the $2$-point functions. Specifically $\Pi_g(p)$, $\Sigma_c(p)$ and 
$\Sigma_q(p)$ are respectively the scalar amplitudes of the gluon polarization 
and the Faddeev-Popov ghost and quark self-energies in the various schemes. The 
particular combination of which of these appears follows from the vertex of 
that MOMi scheme. In deriving the perturbative relations between these two
parameters one has to proceed iteratively order by order in perturbation theory
paying attention to the gauge parameter mapping in the same scheme at the same
time. 

One particular property of the gauge parameter and coupling constant mapping 
between the MOMi and $\MSbar$ schemes is that we can now construct the other 
conversion functions for the wave function renormalizations and the coupling
constant itself. Whilst the latter is not unrelated to (\ref{ccdefs}) we note
that we regard conversion functions as being derived from the explicit forms of
the renormalization constants themselves in the two schemes to be consistent
with other work. Therefore, since we define the coupling constant 
renormalization constant, $Z_g$, by, 
\begin{equation}
g_{\mbox{\footnotesize{o}}} ~=~ \mu^\epsilon Z_g g 
\end{equation}
then the conversion functions are given by
\begin{equation}
C^{\MOMis}_g(a,\alpha) ~=~ \frac{Z_g^{\MOMis}}{Z_g^{\MSbars}} ~~~~,~~~~
C^{\MOMis}_\phi(a,\alpha) ~=~ \frac{Z_\phi^{\MOMis}}{Z_\phi^{\MSbars}}
\end{equation} 
where $\phi$~$\in$~$\{A, \psi, c \}$. We will record the explicit forms of 
these for each of our schemes but note that to determine them we follow the
same iterative procedure as we did in deriving (\ref{almap}). One main benefit
of the conversion functions and the parameter mappings is that we can deduce
the $\beta$-function and anomalous dimensions to {\em three} loops from the
renormalization group. Specifically, \cite{22},  
\begin{equation}
\beta^{\mbox{$\MOMis$}} ( a_{\mbox{$\MOMis$}}, \alpha_{\mbox{$\MOMis$}} ) ~=~
\left[ \beta^{\mbox{$\MSbars$}}( a_{\mbox{$\MSbars$}} ) 
\frac{\partial a_{\mbox{$\MOMis$}}}{\partial a_{\mbox{$\MSbars$}}} \,+\,
\alpha_{\mbox{$\MSbars$}} \gamma^{\mbox{$\MSbars$}}_\alpha 
( a_{\mbox{$\MSbars$}}, \alpha_{\mbox{\footnotesize{$\MSbars$}}} ) 
\frac{\partial a_{\mbox{$\MOMis$}}}{\partial \alpha_{\mbox{$\MSbars$}}} 
\right]_{ \MSbars \rightarrow \MOMis }
\label{betadef}
\end{equation}
and
\begin{eqnarray} 
\gamma_\phi^{\MOMis} ( a_{\MOMis}, \alpha_{\MOMis} )
&=& \!\! \! \left[ \gamma_\phi^{\MSbars} \left(a_{\MSbars}\right) 
+ \beta^{\MSbars}\left(a_{\MSbars}\right) 
\frac{\partial ~}{\partial a_{\MSbars}} \ln C_\phi^{\MOMis} 
\left(a_{\MSbars},\alpha_{\MSbars}\right) \right. \nonumber \\
&& \left. +~ \alpha_{\MSbars} \gamma^{\MSbars}_\alpha 
\left(a_{\MSbars},\alpha_{\MSbars}\right)
\frac{\partial ~}{\partial \alpha_{\MSbars}}
\ln C_\phi^{\MOMis} \left(a_{\MSbars},\alpha_{\MSbars}\right) 
\right]_{ \MSbars \rightarrow \MOMis } \nonumber \\
\label{anomdef}
\end{eqnarray}
where the MOMi $\beta$-functions will depend on the gauge parameter and only be
scheme independent at one loop as these are mass dependent renormalization
schemes, \cite{21}. Though the $\MSbar$ $\beta$-function is independent of
$\alpha$ which is why it has only one argument. The mapping
$\MSbar$~$\rightarrow$~MOMi indicates that the object within the square
brackets is first computed in terms of $\MSbar$ variables and then these 
variables are mapped back to the MOMi scheme variables by inverting 
(\ref{almap}) and those derived from (\ref{ccdefs}). As a check on this 
procedure we have calculated the $\beta$-function and anomalous dimensions to 
two loops for each MOM scheme directly from the two loop renormalization 
constants and verified that the computation contained on the right hand sides 
of (\ref{betadef}) and (\ref{anomdef}) are in agreement at the same order. This
only requires the one loop terms of the conversion functions and therefore this
provides a check on our computer routines designed to do this automatically. 
Finally, in order to ease comparison of our three loop MOMi results with the 
$\MSbar$ scheme in the same notation the numerical expressions for $SU(3)$ for 
the latter scheme are, \cite{1,2,3,4,5}, 
\begin{eqnarray}
\beta^{\mbox{$\MSbars$}}(a) &=& -~ [ 11.0000000 - 0.6666667 \Nf ] a^2
\nonumber \\
&& -~ \left[ 102.0000000 ~-~ 12.6666667 \Nf \right] a^3 \nonumber \\
&& -~ \left[ 1428.5000000 ~-~ 279.6111111 \Nf ~+~ 6.0185185 \Nf^2 \right] 
a^4 ~+~ O(a^5) \nonumber \\
\gamma_A^{\mbox{$\MSbars$}}(a,\alpha) &=& [ 0.6666667 \Nf - 6.5000000
+ 1.5000000 \alpha ] a \nonumber \\
&& -~ \left[ 66.3750000 - 12.3750000 \alpha - 2.2500000 \alpha^2 
- 10.1666667 \Nf \right] a^2 \nonumber \\
&& -~ \left[ 915.9625108 - 165.2479023 \alpha - 33.9291631 \alpha^2
- 5.9062500 \alpha^3 \right. \nonumber \\
&& \left. ~~~~- \left[ 186.8599000 - 9.0000000 \alpha \right] \Nf ~+~ 
7.9629630 \Nf^2 \right] a^3 ~+~ O(a^4) \nonumber \\
\gamma_c^{\mbox{$\MSbars$}}(a,\alpha) &=& [ 0.7500000 \alpha - 2.2500000 ] a 
\nonumber \\
&& -~ \left[ 17.8125000 + 0.5625000 \alpha - 1.2500000 \Nf \right] a^2 
\nonumber \\
&& -~ \left[ 256.2687446 - 2.1729239 \alpha + 0.5114565 \alpha^2
- 1.2656250 \alpha^3 \right. \nonumber \\
&& \left. ~~~~- \left[ 46.2756056 - 3.9375000 \alpha \right] \Nf ~-~
0.9722222 \Nf^2 \right] a^3 ~+~ O(a^4) \nonumber \\
\gamma_\psi^{\mbox{$\MSbars$}}(a,\alpha) &=& 1.3333333 \alpha a \nonumber \\
&& +~ \left[ 22.3333333 + 8.0000000 \alpha + 1.0000000 \alpha^2 ~-~ 
1.3333333 \Nf \right] a^2 \nonumber \\
&& +~ \left[ 528.3243079 + 109.4435121 \alpha + 20.0342561 \alpha^2
+ 3.7500000 \alpha^3 \right. \nonumber \\
&& \left. ~~~~- \left[ 61.1111111 + 8.5000000 \alpha \right] \Nf ~+~
0.7407407 \Nf^2 \right] a^3 ~+~ O(a^4) 
\end{eqnarray}
to the same numerical accuracy. To ease the presentation on the eye we will use
the convention that when the scheme appears on the function on the left hand
side then the variables, such as the coupling constant and gauge parameter, are
the variables in the {\em same} scheme. The exception to this is the conversion
functions, $C_g^{\MOMis}(a,\alpha)$ and $C_\phi^{\MOMis}(a,\alpha)$, where the
arguments are the $\MSbar$ variables. 

\section{Triple gluon vertex.}

We now begin the mundane task of recording our results for each scheme and its 
respective vertices by concentrating on the triple gluon vertex first. As
indicated earlier the full analytic versions of all the results for an
arbitrary gauge, in this and the next two sections, have been included in a 
separate data file. The $\MSbar$ $SU(3)$ numerical values for the amplitudes 
are  
\begin{eqnarray}
\left. \Sigma^{\mbox{\footnotesize{ggg}}}_{(1)}(p,q) \right|_{\MSbars} &=& 
\left. \Sigma^{\mbox{\footnotesize{ggg}}}_{(2)}(p,q) \right|_{\MSbars} ~=~ 
-~ \frac{1}{2} \left. \Sigma^{\mbox{\footnotesize{ggg}}}_{(3)}(p,q) 
\right|_{\MSbars} ~=~ 
-~ \left. \Sigma^{\mbox{\footnotesize{ggg}}}_{(4)}(p,q) \right|_{\MSbars}
\nonumber \\
&=& \frac{1}{2} \left. \Sigma^{\mbox{\footnotesize{ggg}}}_{(5)}(p,q) 
\right|_{\MSbars} ~=~  
-~ \left. \Sigma^{\mbox{\footnotesize{ggg}}}_{(6)}(p,q) \right|_{\MSbars}
\nonumber \\
&=& -~ 1 ~- \left[ 1.1212444 - 3.7618956 \alpha - 1.2890232 \alpha^2
+ 0.1250000 \alpha^3 \right. \nonumber \\
&& \left. ~~~~~~~~~~-~ 0.0417366 \Nf \right] a \nonumber \\
&& +~ \left[ 29.7530676 + 16.4600770 \alpha - 9.7794300 \alpha^2
- 3.2060809 \alpha^3 \right. \nonumber \\
&& \left. ~~~~- 1.6522848 \alpha^4 + 0.2812500 \alpha^5 \right. \nonumber \\
&& \left. ~~~~- [ 11.5677203 - 0.9686976 \alpha - 0.9112399 \alpha^2 
+ 0.4166667 \alpha^3 ] \Nf \right] a^2 \nonumber \\
&& +~ O(a^3) \nonumber \\
\left. \Sigma^{\mbox{\footnotesize{ggg}}}_{(7)}(p,q) \right|_{\MSbars} &=& 
2 \left. \Sigma^{\mbox{\footnotesize{ggg}}}_{(9)}(p,q) \right|_{\MSbars} ~=~
-~ 2 \left. \Sigma^{\mbox{\footnotesize{ggg}}}_{(11)}(p,q) 
\right|_{\MSbars} ~=~ 
-~ \left. \Sigma^{\mbox{\footnotesize{ggg}}}_{(14)}(p,q) \right|_{\MSbars} 
\nonumber \\
&=& \left[ 7.0567163 - 3.3280464 \alpha - 0.5079304 \alpha^2
+ 0.0573179 \alpha^3 - 1.0926858 \Nf \right] a \nonumber \\
&& +~ \left[ 116.0789643 - 13.6830818 \alpha + 0.3484134 \alpha^2
+ 4.7763124 \alpha^3 \right. \nonumber \\
&& \left. ~~~~+ 0.8908609 \alpha^4 - 0.1289652 \alpha^5 \right. \nonumber \\
&& \left. ~~~~- [ 20.2710109 + 1.0153018 \alpha - 0.5745217 \alpha^2 
- 0.1910596 \alpha^3 ] \Nf \right] a^2 \nonumber \\
&& +~ O(a^3) \nonumber \\
\left. \Sigma^{\mbox{\footnotesize{ggg}}}_{(8)}(p,q) \right|_{\MSbars} &=& 
-~ \left. \Sigma^{\mbox{\footnotesize{ggg}}}_{(13)}(p,q) \right|_{\MSbars} 
\nonumber \\
&=& \left[ 7.3683002 - 3.3518377 \alpha - 0.5701159 \alpha^2
+ 0.1926821 \alpha^3 - 1.2130096 \Nf \right] a \nonumber \\
&& +~ \left[ 126.0048710 - 11.8048854 \alpha + 3.7795690 \alpha^2
+ 4.3779190 \alpha^3 \right. \nonumber \\
&& \left. ~~~~+ 1.2887087 \alpha^4 - 0.4335348 \alpha^5 \right. \nonumber \\
&& \left. ~~~~- [ 23.5898191 - 0.0155813 \alpha - 0.9363317 \alpha^2 
- 0.6422738 \alpha^3 ] \Nf \right] a^2 \nonumber \\
&& +~ O(a^3) \nonumber \\
\left. \Sigma^{\mbox{\footnotesize{ggg}}}_{(10)}(p,q) \right|_{\MSbars} &=& 
-~ \left. \Sigma^{\mbox{\footnotesize{ggg}}}_{(12)}(p,q) \right|_{\MSbars} 
\nonumber \\
&=& -\, \left[ 0.3115839 - 0.0237913 \alpha - 0.0621855 \alpha^2
+ 0.1353643 \alpha^3 - 0.1203238 \Nf \right] \! a \nonumber \\
&& -~ \left[ 9.9259067 + 1.8781964 \alpha + 3.4311557 \alpha^2
- 0.3983934 \alpha^3 \right. \nonumber \\
&& \left. ~~~~+ 0.3978478 \alpha^4 - 0.3045696 \alpha^5 \right. \nonumber \\
&& \left. ~~~~- [ 3.3188082 - 1.0308831 \alpha - 0.3618100 \alpha^2 
- 0.4512142 \alpha^3 ] \Nf \right] a^2 \nonumber \\
&& +~ O(a^3) ~. 
\end{eqnarray}
We have indicated the relations between amplitudes of the various projection 
tensor channels. These are consistent with the expectations for the structure 
of the vertex from symmetry given that we have evaluated the vertex function at
the symmetric point. In addition in this context  
\begin{equation}
\left. \Sigma^{\mbox{\footnotesize{ggg}}}_{(7)}(p,q) \right|_{\MSbars} ~=~
\left. \Sigma^{\mbox{\footnotesize{ggg}}}_{(8)}(p,q) \right|_{\MSbars} ~-~ 
\left. \Sigma^{\mbox{\footnotesize{ggg}}}_{(10)}(p,q) \right|_{\MSbars} 
\end{equation}
which we checked was true to two loops prior to the numerical evaluation. In
Table $1$ we give the comparison of our results for the channel $1$ amplitudes
to those of \cite{22}. In this and other comparisons we have used the coupling 
constant convention of the presentation in \cite{22} where the series was in 
powers of $\alpha_s$~$=$~$g^2/(4\pi^2)$. This is to retain the error estimates 
given in \cite{22}. It is clear that our evaluation is comfortably close to the
approximations of \cite{22}. 

{\begin{table}[ht]
\begin{center}
\begin{tabular}{|c||r|r|r|}
\hline
$\left. \Sigma^{\mbox{\footnotesize{ggg}}}_{(1),2}(p,q) \right|_{\MSbars}$ 
& $C_A^2$ & $C_A T_F \Nf$ & $C_F T_F \Nf$ \\
\hline
Ref \cite{22} & $0.22(4)$ & $-$ $0.65(7)$ & $0.408(10)$ \\ 
This paper & $0.2066185$ & $-$ $0.6620808$ & $0.4052081$ \\ 
\hline
\end{tabular}
\end{center}
\begin{center}
{Table $1$. Comparison of channel $1$ two loop Landau gauge amplitude with 
\cite{22} by colour factor.}  
\end{center}
\end{table}}

The MOMggg scheme amplitudes satisfy the same relations and in particular we 
have 
\begin{eqnarray}
\left. \Sigma^{\mbox{\footnotesize{ggg}}}_{(1)}(p,q) \right|_{\MOMgggs} &=& 
\left. \Sigma^{\mbox{\footnotesize{ggg}}}_{(2)}(p,q) \right|_{\MOMgggs} ~=~ 
-~ \frac{1}{2} \left. \Sigma^{\mbox{\footnotesize{ggg}}}_{(3)}(p,q) 
\right|_{\MOMgggs} \nonumber \\
&=& -~ \left. \Sigma^{\mbox{\footnotesize{ggg}}}_{(4)}(p,q)
\right|_{\MOMgggs} ~=~ 
\frac{1}{2} \left. \Sigma^{\mbox{\footnotesize{ggg}}}_{(5)}(p,q) 
\right|_{\MOMgggs} \nonumber \\
&=& -~ \left. \Sigma^{\mbox{\footnotesize{ggg}}}_{(6)}(p,q) 
\right|_{\MOMgggs} ~=~ -~ 1 ~+~ O(a^3) \nonumber \\
\left. \Sigma^{\mbox{\footnotesize{ggg}}}_{(7)}(p,q) \right|_{\MOMgggs} &=& 
2 \left. \Sigma^{\mbox{\footnotesize{ggg}}}_{(9)}(p,q) \right|_{\MOMgggs} ~=~
-~ 2 \left. \Sigma^{\mbox{\footnotesize{ggg}}}_{(11)}(p,q) 
\right|_{\MOMgggs} \nonumber \\
&=& -~ \left. \Sigma^{\mbox{\footnotesize{ggg}}}_{(14)}(p,q) \right|_{\MOMgggs} 
\nonumber \\
&=& \left[ 7.0567163 - 3.3280464 \alpha - 0.5079304 \alpha^2
+ 0.0573179 \alpha^3 \right. \nonumber \\
&& \left. ~- 1.0926858 \Nf \right] a \nonumber \\
&& -~ \left[ 78.7833165 - 99.1996625 \alpha + 10.0012225 \alpha^2
+ 10.9109237 \alpha^3 \right. \nonumber \\
&& \left. ~~~~- 1.2024953 \alpha^4 - 0.2831609 \alpha^5 + 0.0214942 \alpha^6 
\right. \nonumber \\
&& \left. ~~~~- [ 34.3080792 - 16.2422875 \alpha - 1.8203921 \alpha^2 
+ 0.6079935 \alpha^3 ] \Nf \right. \nonumber \\
&& \left. ~~~~+ 3.7791007 \Nf^2 \right] a^2 ~+~ O(a^3) \nonumber \\
\left. \Sigma^{\mbox{\footnotesize{ggg}}}_{(8)}(p,q) \right|_{\MOMgggs} &=& 
-~ \left. \Sigma^{\mbox{\footnotesize{ggg}}}_{(13)}(p,q) \right|_{\MOMgggs} 
\nonumber \\
&=& \left[ 7.3683002 - 3.3518377 \alpha - 0.5701159 \alpha^2
+ 0.1926821 \alpha^3 \right. \nonumber \\
&& \left. ~- 1.2130096 \Nf \right] a \nonumber \\
&& -~ \left[ 77.4614044 - 103.6568232 \alpha + 5.5514994 \alpha^2
+ 12.5463340 \alpha^3 \right. \nonumber \\
&& \left. ~~~~- 2.9431069 \alpha^4 - 0.5253739 \alpha^5 + 0.07225580 \alpha^6
\right. \nonumber \\
&& \left. ~~~~- [ 35.3894861 - 16.0837324 \alpha - 1.7300353 \alpha^2 
+ 1.1212780 \alpha^3 ] \Nf \right. \nonumber \\
&& \left. ~~~~+ 4.1952457 \Nf^2 \right] a^2 ~+~ O(a^3) \nonumber \\
\left. \Sigma^{\mbox{\footnotesize{ggg}}}_{(10)}(p,q) \right|_{\MOMgggs} &=& 
-~ \left. \Sigma^{\mbox{\footnotesize{ggg}}}_{(12)}(p,q) \right|_{\MOMgggs} 
\nonumber \\
&=& -\, \left[ 0.3115839 - 0.0237913 \alpha - 0.0621855 \alpha^2
+ 0.1353643 \alpha^3 \right. \nonumber \\
&& \left. ~~~~- 0.1203238 \Nf \right] a \nonumber \\
&& -~ \left[ 1.3219120 + 4.4571607 \alpha + 4.44972304\alpha^2
- 1.6354103 \alpha^3 \right. \nonumber \\
&& \left. ~~~~+ 1.7406116 \alpha^4 + 0.2422130 \alpha^5 - 0.0507616 \alpha^6
\right. \nonumber \\
&& \left. ~~~~+ [ 1.0814069 + 0.1585516 \alpha + 0.0903568 \alpha^2 
+ 0.5132845 \alpha^3 ] \Nf \right. \nonumber \\ 
&& \left. ~~~~- 0.4161497 \Nf^2 \right] a^2 ~+~ O(a^3) 
\end{eqnarray}
with the corresponding relation  
\begin{equation}
\left. \Sigma^{\mbox{\footnotesize{ggg}}}_{(7)}(p,q) \right|_{\MOMgggs} ~=~
\left. \Sigma^{\mbox{\footnotesize{ggg}}}_{(8)}(p,q) \right|_{\MOMgggs} ~-~ 
\left. \Sigma^{\mbox{\footnotesize{ggg}}}_{(10)}(p,q) \right|_{\MOMgggs} 
\end{equation}
also being true to two loops analytically. Given the nature of the MOMggg 
scheme the relations for the amplitudes of channels $1$ to $6$ demonstrate that
our renormalization is consistent and that our projection has been implemented 
consistently within our {\sc Form} programmes as well as being a check on our 
{\sc Reduze} database. Next we record the four conversion functions for the 
wave function and coupling constant renormalizations in the MOMggg scheme 
relative to the $\MSbar$ scheme. These are 
\begin{eqnarray}
C_g^{\MOMgggs} (a,\alpha) &=& 1 ~-~ \left[ 13.2462444
- 1.5118956 \alpha - 0.1640232 \alpha^2 \right. \nonumber \\
&& \left. ~~~~~~~~+ 0.1250000 \alpha^3 - 1.7084032 \Nf \right] a \nonumber \\
&& - \left[ 217.0368707 + 36.7247782 \alpha + 7.0535877 \alpha^2
- 1.1557619 \alpha^3 \right. \nonumber \\
&& \left. ~~~~+ 1.0453915 \alpha^4 - 0.0996192 \alpha^5 
- 0.0156250 \alpha^6 \right. \nonumber \\
&& \left. ~~~~- [ 33.1527255 + 3.7086335 \alpha - 0.0047429 \alpha^2 
- 0.6354341 \alpha^3 ] \Nf \right. \nonumber \\
&& \left. ~~~~- 0.5342658 \Nf^2 \right] a^2 ~+~ O(a^3) \nonumber \\
C_A^{\MOMgggs}(a,\alpha) &=& 1 ~+~ \left[ 8.0833333 
+ 1.5000000 \alpha + 0.7500000 \alpha^2 - 1.1111111 \Nf \right] a \nonumber \\
&& + \left[ 256.1034914 + 31.4182743 \alpha + 8.4375000 \alpha^2
+ 2.8125000 \alpha^3 \right] \nonumber \\
&& \left. ~~~~- [ 53.9129277 + 1.6666667 \alpha ] \Nf ~+~ 1.2345679 \Nf^2 
\right] a^2 ~+~ O(a^3) \nonumber \\
C_c^{\MOMgggs}(a,\alpha) &=& 1 ~+~ 3.0000000 a \nonumber \\
&& +~ \left[ 80.9357699 + 3.0725670 \alpha + 1.3465290 \alpha^2 - 5.9375000 \Nf
\right] a^2 \nonumber \\
&& +~ O(a^3) \nonumber \\
C_\psi^{\MOMgggs}(a,\alpha) &=& 1 ~-~ 1.3333333 \alpha a \nonumber \\
&& - \left[ 25.4642061 + 11.5753172 \alpha + 2.7222222 \alpha^2 
- 2.3333333 \Nf \right] a^2 \nonumber \\
&& +~ O(a^3) 
\end{eqnarray}
where we note that the expressions for all bar $C_g^{\MOMgggs} (a,\alpha)$ are 
formally the same as those for the RI${}^\prime$ scheme. By definition 
$C_g^{\RIps} (a,\alpha)$ is unity. From $C_g^{\MOMgggs} (a,\alpha)$ we can 
deduce the explicit form of the coupling constant mapping. Though as the full 
form is cumbersome we record the analytic version for the Landau gauge which is
\begin{eqnarray}
a_{\MOMgggs} &=& a_{\MSbars} 
+ \left[ \left[ 69 \psi^\prime(\third) - 46 \pi^2 + 1188 \right] C_A
+ \left[ 128 \pi^2 - 192 \psi^\prime(\third) - 432 \right] T_F \Nf \right] 
\frac{a_{\MSbars}^2}{162} \nonumber \\ 
&& +~ \left[ \left[ 
19044 (\psi^\prime(\third))^2
- 25392 \pi^2 \psi^\prime(\third) 
- 6938784 \psi^\prime(\third)
- 100602 \psi^{\prime\prime\prime}(\third) 
\right. \right. \nonumber \\
&& \left. \left. ~~~~~
-~ 72643392 s_2(\pisix)
+ 145286784 s_2(\pitwo)
+ 121072320 s_3(\pisix)
- 96857856 s_3(\pitwo)
\right. \right. \nonumber \\
&& \left. \left. ~~~~~
+~ 276736 \pi^4 
+ 4625856 \pi^2 
- 113724 \Sigma 
+ 8301852 \zeta(3) 
+ 40126833
\right. \right. \nonumber \\
&& \left. \left. ~~~~~
-~ 504468 \frac{\ln^2(3) \pi}{\sqrt{3}} 
+ 6053616 \frac{\ln(3) \pi}{\sqrt{3}}
+ 541836 \frac{\pi^3}{\sqrt{3}}
\right] C_A^2 \right. \nonumber \\
&& ~~~~~+ \left. \left[ 
141312 \pi^2 \psi^\prime(\third) 
- 105984 (\psi^\prime(\third))^2
- 2960064 \psi^\prime(\third)
+ 33592320 s_2(\pisix)
\right. \right. \nonumber \\
&& \left. \left. ~~~~~~~~~
-~ 67184640 s_2(\pitwo)
- 55987200 s_3(\pisix)
+ 44789760 s_3(\pitwo)
- 47104 \pi^4 
\right. \right. \nonumber \\
&& \left. \left. ~~~~~~~~~
+~ 1973376 \pi^2 
+ 2239488 \Sigma 
- 8957952 \zeta(3) 
- 26695008
\right. \right. \nonumber \\
&& \left. \left. ~~~~~~~~~
+~ 233280 \frac{\ln^2(3) \pi}{\sqrt{3}} 
- 2799360 \frac{\ln(3) \pi}{\sqrt{3}}
- 250560 \frac{\pi^3}{\sqrt{3}}
\right] C_A T_F \Nf \right. \nonumber \\
&& ~~~~~+ \left. \left[ 
124416 \psi^{\prime\prime\prime}(\third) 
- 1492992 \psi^\prime(\third)
- 331776 \pi^4 
+ 995328 \pi^2 
\right. \right. \nonumber \\
&& \left. \left. ~~~~~~~~~
-~ 4478976 \Sigma
+ 6718464 \zeta(3)
- 7138368 
\right] C_F T_F \Nf \right. \nonumber \\ 
&& ~~~~~+ \left. \left[ 
147456 (\psi^\prime(\third))^2
- 196608 \pi^2 \psi^\prime(\third) 
+ 2322432 \psi^\prime(\third)
+ 65536 \pi^4 
\right. \right. \nonumber \\
&& \left. \left. ~~~~~~~~~
-~ 1548288 \pi^2 
+ 2923776 
\right] T_F^2 \Nf^2 \right] \frac{a_{\MSbars}^3}{419904} ~+~ 
O\left( a_{\MSbars}^4 \right) ~.
\end{eqnarray}
By contrast the full gauge dependent version in numerical form for $SU(3)$ is 
\begin{eqnarray}
a_{\MOMgggs} &=& a_{\MSbars} \nonumber \\
&& +~ \left[ 26.4924889 - 3.0237913 \alpha_{\MSbars} 
- 0.3280464 \alpha_{\MSbars}^2 \right. \nonumber \\
&& \left. ~~~~+ 0.2500000 \alpha_{\MSbars}^3 - 3.4168064 \Nf \right] 
a_{\MSbars}^2 \nonumber \\
&& +~ \left[ 960.4627167 - 46.7120794 \alpha_{\MSbars} 
+ 7.9285132 \alpha_{\MSbars}^2 + 9.1110752 \alpha_{\MSbars}^3 
\right. \nonumber \\
&& \left. ~~~~+ 1.0375721 \alpha_{\MSbars}^4 - 0.3222558 \alpha_{\MSbars}^5 
+ 0.0156250 \alpha_{\MSbars}^6 \right. \nonumber \\
&& \left. ~~~~- \left[ 202.0850109 - 8.0802973 \alpha_{\MSbars} 
- 1.6907923 \alpha_{\MSbars}^2 \right. \right. \nonumber \\
&& \left. \left. ~~~~~~~+ 0.0104341 \alpha_{\MSbars}^3 \right] \Nf 
+ 7.6873930 \Nf^2 \right] a_{\MSbars}^3 ~+~ O\left( a_{\MSbars}^4 \right) ~. 
\end{eqnarray}
Equipped with these we have computed the three loop renormalization group
functions for the MOMggg scheme. Again given space considerations we record the 
Landau gauge expressions for each case. They are 
\begin{eqnarray}
\beta^{\mbox{$\MOMgggs$}}(a,0) &=& -~ \left[ \frac{11}{3} C_A - \frac{4}{3} 
T_F \Nf \right] a^2 ~-~ \left[ \frac{34}{3} C_A^2 - 4 C_F T_F \Nf 
- \frac{20}{3} C_A T_F \Nf \right] a^3 \nonumber \\
&& +~ \left[ \left[ 
209484 (\psi^\prime(\third))^2
- 279312 \pi^2 \psi^\prime(\third) 
+ 37087200 \psi^\prime(\third)
\right. \right. \nonumber \\
&& \left. \left. ~~~~~
+~ 368874 \psi^{\prime\prime\prime}(\third) 
+ 266359104 s_2(\pisix)
- 532718208 s_2(\pitwo)
\right. \right. \nonumber \\
&& \left. \left. ~~~~~
-~ 443931840 s_3(\pisix)
+ 355145472 s_3(\pitwo)
- 890560 \pi^4 
\right. \right. \nonumber \\
&& \left. \left. ~~~~~
-~ 24724800 \pi^2 
+ 416988 \Sigma 
- 30440124 \zeta(3) 
- 51650217
\right. \right. \nonumber \\
&& \left. \left. ~~~~~
+~ 1849716 \frac{\ln^2(3) \pi}{\sqrt{3}} 
- 22196592 \frac{\ln(3) \pi}{\sqrt{3}}
- 1986732 \frac{\pi^3}{\sqrt{3}}
\right] C_A^3 \right. \nonumber \\
&& ~~~~~+ \left. \left[ 
1656000 \pi^2 \psi^\prime(\third) 
- 1242000 (\psi^\prime(\third))^2
- 38988864 \psi^\prime(\third)
\right. \right. \nonumber \\
&& \left. \left. ~~~~~~~~~
-~ 134136 \psi^{\prime\prime\prime}(\third) 
- 220029696 s_2(\pisix)
+ 440059392 s_2(\pitwo)
\right. \right. \nonumber \\
&& \left. \left. ~~~~~~~~~
+~ 366716160 s_3(\pisix)
- 293372928 s_3(\pitwo)
- 194304 \pi^4 
\right. \right. \nonumber \\
&& \left. \left. ~~~~~~~~~
+~ 25992576 \pi^2 
- 8363088 \Sigma 
+ 43914960 \zeta(3) 
+ 49845132
\right. \right. \nonumber \\
&& \left. \left. ~~~~~~~~~
-~ 1527984 \frac{\ln^2(3) \pi}{\sqrt{3}} 
+ 18335808 \frac{\ln(3) \pi}{\sqrt{3}}
+ 1641168 \frac{\pi^3}{\sqrt{3}}
\right] C_A^2 T_F \Nf \right. \nonumber \\
&& ~~~~~+ \left. \left[ 
2045952 (\psi^\prime(\third))^2
- 2727936 \pi^2 \psi^\prime(\third) 
+ 11591424 \psi^\prime(\third)
\right. \right. \nonumber \\
&& \left. \left. ~~~~~~~~~
+~ 44789760 s_2(\pisix)
- 89579520 s_2(\pitwo)
- 74649600 s_3(\pisix)
\right. \right. \nonumber \\
&& \left. \left. ~~~~~~~~~
+~ 59719680 s_3(\pitwo)
+ 909312 \pi^4 
- 7727616 \pi^2 
\right. \right. \nonumber \\
&& \left. \left. ~~~~~~~~~
+~ 2985984 \Sigma 
- 11943936 \zeta(3) 
- 8460288
\right. \right. \nonumber \\
&& \left. \left. ~~~~~~~~~
+~ 311040 \frac{\ln^2(3) \pi}{\sqrt{3}} 
- 3732480 \frac{\ln(3) \pi}{\sqrt{3}}
- 334080 \frac{\pi^3}{\sqrt{3}}
\right] C_A T_F^2 \Nf^2 \right. \nonumber \\
&& ~~~~~+ \left. \left[ 
786432 \pi^2 \psi^\prime(\third) 
- 589824 (\psi^\prime(\third))^2
- 442368 \psi^\prime(\third)
\right. \right. \nonumber \\
&& \left. \left. ~~~~~~~~~
-~ 262144 \pi^4 
+ 294912 \pi^2 
- 82944
\right] T_F^3 \Nf^3 \right. \nonumber \\
&& ~~~~~+ \left. \left[ 
4758912 \psi^\prime(\third)
- 456192 \psi^{\prime\prime\prime}(\third) 
+ 1216512 \pi^4 
- 3172608 \pi^2 
\right. \right. \nonumber \\
&& \left. \left. ~~~~~~~~~
+~ 16422912 \Sigma 
- 24634368 \zeta(3) 
+ 23421312
\right] C_A C_F T_F \Nf \right. \nonumber \\
&& ~~~~~+ \left. \left[ 
165888 \psi^{\prime\prime\prime}(\third) 
- 442368 \pi^4 
- 5971968 \Sigma 
+ 8957952 \zeta(3) 
\right. \right. \nonumber \\
&& \left. \left. ~~~~~~~~~
-~ 7091712
\right] C_F T_F^2 \Nf^2 ~-~ 839808 C_F^2 T_F \Nf \right] \frac{a^4}{419904}
\nonumber \\
&& +~ O(a^5)
\end{eqnarray}
\begin{eqnarray}
\gamma_A^{\mbox{$\MOMgggs$}}(a,0) &=& -~ \left[ 13 C_A - 8 T_F \Nf \right] 
\frac{a}{6} \nonumber \\
&& + \left[ \left[ 1794 \psi^\prime(\third) - 1196 \pi^2 - 2655 \right] C_A^2
+ 7776 C_F T_F \Nf \right. \nonumber \\
&& \left. ~~~+ \left[ 4064 \pi^2 - 6096 \psi^\prime(\third) + 2304 \right] 
C_A T_F \Nf \right. \nonumber \\
&& \left. ~~~+ \left[ 3072 \psi^\prime(\third) - 2048 \pi^2 + 1152 \right]
T_F^2 \Nf^2 \right] \frac{a^2}{1944} \nonumber \\ 
&& +~ \left[ \left[ 
2310672 \pi^2 \psi^\prime(\third) 
- 1733004 (\psi^\prime(\third))^2
- 121373424 \psi^\prime(\third)
\right. \right. \nonumber \\
&& \left. \left. ~~~~~
-~ 1307826 \psi^{\prime\prime\prime}(\third) 
- 944364096 s_2(\pisix)
+ 1888728192 s_2(\pitwo)
\right. \right. \nonumber \\
&& \left. \left. ~~~~~
+~ 1573940160 s_3(\pisix)
- 1259152128 s_3(\pitwo)
+ 2717312 \pi^4 
\right. \right. \nonumber \\
&& \left. \left. ~~~~~
+~ 80915616 \pi^2 
- 1478412 \Sigma 
+ 164768580 \zeta(3) 
- 117299583
\right. \right. \nonumber \\
&& \left. \left. ~~~~~
-~ 6558084 \frac{\ln^2(3) \pi}{\sqrt{3}} 
+ 78697008 \frac{\ln(3) \pi}{\sqrt{3}}
+ 7043868 \frac{\pi^3}{\sqrt{3}}
\right] C_A^3 \right. \nonumber \\
&& ~~~~~+ \left. \left[ 
1071108 (\psi^\prime(\third))^2
- 14281344 \pi^2 \psi^\prime(\third) 
+ 134602560 \psi^\prime(\third)
\right. \right. \nonumber \\
&& \left. \left. ~~~~~~~~~
+~ 804816 \psi^{\prime\prime\prime}(\third) 
+ 1017847296 s_2(\pisix)
- 2035694592 s_2(\pitwo)
\right. \right. \nonumber \\
&& \left. \left. ~~~~~~~~~
-~ 1696412160 s_3(\pisix)
+ 1357129728 s_3(\pitwo)
+ 2614272 \pi^4 
\right. \right. \nonumber \\
&& \left. \left. ~~~~~~~~~
-~ 89735040 \pi^2 
+ 30023136 \Sigma 
- 100567008 \zeta(3) 
+ 80188056
\right. \right. \nonumber \\
&& \left. \left. ~~~~~~~~~
+~ 7068384 \frac{\ln^2(3) \pi}{\sqrt{3}} 
- 84820608 \frac{\ln(3) \pi}{\sqrt{3}}
- 7591968 \frac{\pi^3}{\sqrt{3}}
\right] C_A^2 T_F \Nf \right. \nonumber \\
&& ~~~~~+ \left. \left[ 
25804800 \pi^2 \psi^\prime(\third) 
- 19353600 (\psi^\prime(\third))^2
- 40849920 \psi^\prime(\third)
\right. \right. \nonumber \\
&& \left. \left. ~~~~~~~~~
-~ 268738560 s_2(\pisix)
+ 537477120 s_2(\pitwo)
\right. \right. \nonumber \\
&& \left. \left. ~~~~~~~~~
+~ 447897600 s_3(\pisix)
- 358318080 s_3(\pitwo)
- 8601600 \pi^4 
\right. \right. \nonumber \\
&& \left. \left. ~~~~~~~~~
+~ 27233280 \pi^2 
- 17915904 \Sigma 
+ 17915904 \zeta(3) 
+ 2052864
\right. \right. \nonumber \\
&& \left. \left. ~~~~~~~~~
-~ 1866240 \frac{\ln^2(3) \pi}{\sqrt{3}} 
+ 22394880 \frac{\ln(3) \pi}{\sqrt{3}}
+ 2004480 \frac{\pi^3}{\sqrt{3}}
\right] C_A T_F^2 \Nf^2 \right. \nonumber \\
&& ~~~~~+ \left. \left[ 
8257536 (\psi^\prime(\third))^2
- 11010048 \pi^2 \psi^\prime(\third) 
+ 6193152 \psi^\prime(\third)
+ 3670016 \pi^4 
\right. \right. \nonumber \\
&& \left. \left. ~~~~~~~~~
-~ 4128768 \pi^2 
+ 1161216 
\right] T_F^3 \Nf^3 \right. \nonumber \\
&& ~~~~~+ \left. \left[ 
1617408 \psi^{\prime\prime\prime}(\third) 
- 27993600 \psi^\prime(\third)
- 4313088 \pi^4 
+ 18662400 \pi^2 
\right. \right. \nonumber \\
&& \left. \left. ~~~~~~~~~
-~ 58226688 \Sigma 
- 147806208 \zeta(3) 
+ 125971200
\right] C_A C_F T_F \Nf \right. \nonumber \\
&& ~~~~~+ \left. \left[ 
35831808 \psi^\prime(\third)
- 995328 \psi^{\prime\prime\prime}(\third) 
+ 2654208 \pi^4 
- 23887872 \pi^2 
\right. \right. \nonumber \\
&& \left. \left. ~~~~~~~~~
+~ 35831808 \Sigma 
+ 53747712 \zeta(3) 
- 47029248
\right] C_F T_F^2 \Nf^2 \right. \nonumber \\
&& \left. ~~~~~-~ 5038848 C_F^2 T_F \Nf \right] \frac{a^3}{2519424} ~+~ O(a^4)
\end{eqnarray}
\begin{eqnarray}
\gamma_c^{\mbox{$\MOMgggs$}}(a,0) &=& -~ \frac{3}{4} C_A a \nonumber \\
&& +~ \left[ \left[ 138 \psi^\prime(\third) - 92 \pi^2 - 63 \right] C_A
\right. \nonumber \\
&& \left. ~~~~+ \left[ 256 \pi^2 - 384 \psi^\prime(\third) + 72 \right] 
T_F \Nf \right] \frac{C_A a^2}{432} \nonumber \\ 
&& +~ \left[ \left[ 
177744 \pi^2 \psi^\prime(\third) 
- 133308 (\psi^\prime(\third))^2
- 9492336 \psi^\prime(\third)
\right. \right. \nonumber \\
&& \left. \left. ~~~~~
-~ 100602 \psi^{\prime\prime\prime}(\third) 
- 72643392 s_2(\pisix)
+ 145286784 s_2(\pitwo)
\right. \right. \nonumber \\
&& \left. \left. ~~~~~
+~ 121072320 s_3(\pisix)
- 96857856 s_3(\pitwo)
+ 209024 \pi^4 
\right. \right. \nonumber \\
&& \left. \left. ~~~~~
+~ 6328224 \pi^2 
- 113724 \Sigma 
+ 11993508 \zeta(3) 
- 9641835
\right. \right. \nonumber \\
&& \left. \left. ~~~~~
-~ 504468 \frac{\ln^2(3) \pi}{\sqrt{3}} 
+ 6053616 \frac{\ln(3) \pi}{\sqrt{3}}
+ 541836 \frac{\pi^3}{\sqrt{3}}
\right] C_A^2 \right. \nonumber \\
&& ~~~~~+ \left. \left[ 
741888 (\psi^\prime(\third))^2
- 989184 \pi^2 \psi^\prime(\third) 
+ 5019840 \psi^\prime(\third)
\right. \right. \nonumber \\
&& \left. \left. ~~~~~~~~~
+~ 33592320 s_2(\pisix)
- 67184640 s_2(\pitwo)
\right. \right. \nonumber \\
&& \left. \left. ~~~~~~~~~
-~ 55987200 s_3(\pisix)
+ 44789760 s_3(\pitwo)
+ 329728 \pi^4 
\right. \right. \nonumber \\
&& \left. \left. ~~~~~~~~~
-~ 3346560 \pi^2 
+ 2239488 \Sigma 
- 5318784 \zeta(3) 
+ 6350400
\right. \right. \nonumber \\
&& \left. \left. ~~~~~~~~~
+~ 233280 \frac{\ln^2(3) \pi}{\sqrt{3}} 
- 2799360 \frac{\ln(3) \pi}{\sqrt{3}}
- 250560 \frac{\pi^3}{\sqrt{3}}
\right] C_A T_F \Nf \right. \nonumber \\
&& ~~~~~+ \left. \left[ 
1376256 \pi^2 \psi^\prime(\third) 
- 1032192 (\psi^\prime(\third))^2
- 110592 \psi^\prime(\third)
\right. \right. \nonumber \\
&& \left. \left. ~~~~~~~~~
-~ 458752 \pi^4 
+ 73728 \pi^2 
- 1762560
\right] T_F^2 \Nf^2 \right. \nonumber \\
&& ~~~~~+ \left. \left[ 
124416 \psi^\prime(\third)
- 1492992 \psi^\prime(\third)
-  331776 \pi^4 
\right. \right. \nonumber \\
&& \left. \left. ~~~~~~~~~
+~ 995328 \pi^2 
- 4478976 \Sigma
+ 1399680
\right] C_F T_F \Nf \right] \frac{C_A a^3}{559872} \nonumber \\
&& +~ O(a^4) 
\end{eqnarray}
and 
\begin{eqnarray}
\gamma_\psi^{\mbox{$\MOMgggs$}}(a,0) &=& \left[ 25 C_A - 6 C_F 
- 8 T_F \Nf \right] \frac{C_F a^2}{4} \nonumber \\
&& +~ \left[ \left[ 4600 \pi^2 - 6900 \psi^\prime(\third)
- 39690 \zeta(3) + 61011 \right] C_A^2 \right. \nonumber \\
&& \left. ~~~~+ \left[ 1656 \psi^\prime(\third) - 1104 \pi^2 
+ 15552 \zeta(3) - 23760 \right] C_A C_F \right. \nonumber \\
&& \left. ~~~~+ \left[ 21408 \psi^\prime(\third) - 14272 \pi^2 
+ 10368 \zeta(3) - 28800 \right] C_A T_F \Nf \right. \nonumber \\
&& \left. ~~~~+ \left[ 3072 \pi^2 - 4608 \psi^\prime(\third) 
- 4320 \right] C_F T_F \Nf ~+~ 1944 C_F^2 \right. \nonumber \\
&& \left. ~~~~+ \left[ 4096 \pi^2 - 6144 \psi^\prime(\third) 
+ 1152 \right] T_F^2 \Nf^2 \right] \frac{C_F a^2}{1296} 
\nonumber \\
&& +~ O(a^4) ~.
\end{eqnarray}
That for the quark anomalous dimension is relatively compact as there is no one
loop term in the Landau gauge. The explicit gauge dependence is given in the
numerical evaluations for $SU(3)$ which are 
\begin{eqnarray}
\beta^{\mbox{$\MOMgggs$}}(a,\alpha) &=& -~ [ 11.0000000 - 0.6666667 \Nf ] a^2
\nonumber \\
&& -~ \left[ 102.0000000 + 19.6546434 \alpha - 0.2710840 \alpha^2
- 5.8591391 \alpha^3 \right. \nonumber \\
&& \left. ~~~~+ 1.1250000 \alpha^4 \right. \nonumber \\
&& \left. ~~~~- \left[ 12.6666667 + 2.0158609 \alpha + 0.4373952 \alpha^2
- 0.5000000 \alpha^3 \right] \Nf \right] a^3 \nonumber \\
&& -~ \left[ 1570.9843804 + 658.0709292 \alpha + 269.2238338 \alpha^2
+ 43.0029610 \alpha^3 \right. \nonumber \\
&& \left. ~~~~- 99.2797189 \alpha^4 + 14.8550247 \alpha^5
+ 5.3345924 \alpha^6 - 0.7031250 \alpha^7 \right. \nonumber \\
&& \left. ~~~~+ \left[ 0.5659290 - 43.2393672 \alpha - 22.7471960 \alpha^2
- 19.8709555 \alpha^3 \right. \right. \nonumber \\
&& \left. \left. ~~~~~~~~+ 14.8347569 \alpha^4 + 0.9764184 \alpha^5
- 0.2812500 \alpha^6 \right] \Nf \right. \nonumber \\
&& \left. ~~~~- \left[ 67.0895364 + 4.6479610 \alpha + 0.8898051 \alpha^2
- 2.3056953 \alpha^3 \right] \Nf^2 \right. \nonumber \\
&& \left. ~~~~+ 2.6581155 \Nf^3 \right] a^4 ~+~ O(a^5) \nonumber \\
\gamma_A^{\mbox{$\MOMgggs$}}(a,\alpha) &=& [ 0.6666667 \Nf - 6.5000000
+ 1.5000000 \alpha ] a \nonumber \\
&& +~ \left[ 16.9095110 - 41.6433767 \alpha + 6.1533855 \alpha^2
+ 0.9920696 \alpha^3 \right. \nonumber \\
&& \left. ~~~~- 0.3750000 \alpha^4 \right. \nonumber \\
&& \left. ~~~~- \left[ 12.0931233 - 5.4744039 \alpha + 0.2813024 \alpha^2
+ 0.1666667 \alpha^3 \right] \Nf \right. \nonumber \\
&& \left. ~~~~+ 1.5371302 \Nf^2 \right] a^2 \nonumber \\
&& -~ \left[ 1308.9386744 - 647.9260677 \alpha + 376.2301295 \alpha^2
+ 6.3971133 \alpha^3 \right. \nonumber \\
&& \left. ~~~~- 33.0162468 \alpha^4 + 7.3253130 \alpha^5
+ 1.0008734 \alpha^6 - 0.1640625 \alpha^7 \right. \nonumber \\
&& \left. ~~~~- \left[ 491.4309500 - 302.3530495 \alpha + 52.3029150 \alpha^2
+ 6.3604344 \alpha^3 \right. \right. \nonumber \\
&& \left. \left. ~~~~~~~~- 6.7153148 \alpha^4 - 0.1288604 \alpha^5
+ 0.0729167 \alpha^6 \right] \Nf \right. \nonumber \\
&& \left. ~~~~+ \left[ 74.9190172 - 29.3999306 \alpha + 1.4158902 \alpha^2
+ 1.3449889 \alpha^3 \right] \Nf^2 \right. \nonumber \\
&& \left. ~~~~- 6.2022694 \Nf^3 \right] a^3 ~+~ O(a^4) \nonumber \\
\gamma_c^{\mbox{$\MOMgggs$}}(a,\alpha) &=& [ 0.7500000 \alpha - 2.2500000 ] a 
\nonumber \\
&& +~ \left[ 8.7955100 - 21.1728971 \alpha + 2.6547391 \alpha^2
+ 1.3710348 \alpha^3 \right. \nonumber \\
&& \left. ~~~~- 0.1875000 \alpha^4 ~-~ \left[ 4.4378145 
- 1.7292715 \alpha \right] \Nf \right] a^2 \nonumber \\
&& -~ \left[ 548.8492387 - 436.6720556 \alpha + 199.2938032 \alpha^2
+ 32.7086138 \alpha^3 \right. \nonumber \\
&& \left. ~~~~- 30.0123943 \alpha^4 + 1.4303427 \alpha^5
+ 0.9535617 \alpha^6 - 0.0820312 \alpha^7 \right. \nonumber \\
&& \left. ~~~~- \left[ 157.4669179 - 127.5457558 \alpha + 20.0522192 \alpha^2
+ 7.7592757 \alpha^3 \right. \right. \nonumber \\
&& \left. \left. ~~~~~~~~- 1.5131126 \alpha^4 \right] \Nf ~+~ \left[ 
19.9741163 - 6.9775531 \alpha \right] \Nf^2 \right] a^3 ~+~ O(a^4) \nonumber \\
\gamma_\psi^{\mbox{$\MOMgggs$}}(a,\alpha) &=& 1.3333333 \alpha a \nonumber \\
&& +~ \left[ 22.3333333 - 10.5455407 \alpha + 9.0317217 \alpha^2
+ 1.4373952 \alpha^3 \right. \nonumber \\
&& \left. ~~~~- 0.3333333 \alpha^4 ~-~ \left[ 1.3333333 
- 3.0742604 \alpha \right] \Nf \right] a^2 \nonumber \\
&& -~ \left[ 94.7943290 - 204.1998798 \alpha + 218.8404110 \alpha^2
- 30.4216658 \alpha^3 \right. \nonumber \\
&& \left. ~~~~- 34.4073860 \alpha^4 + 6.3159939 \alpha^5
+ 1.2577208 \alpha^6 - 0.1458333 \alpha^7 \right. \nonumber \\
&& \left. ~~~~- \left[ 76.8672720 - 80.5601965 \alpha + 53.0718118 \alpha^2
+ 7.0576676 \alpha^3 \right. \right. \nonumber \\
&& \left. \left. ~~~~~~~~- 2.6899779 \alpha^4 \right] \Nf ~+~ \left[ 
5.2596320 - 12.4045389 \alpha \right] \Nf^2 \right] a^3 \nonumber \\
&& +~ O(a^4) ~. 
\end{eqnarray}
We recall, \cite{22}, that unlike the $\MSbar$ scheme the three loop term of 
the MOMggg $\beta$-function is cubic in $\Nf$ and not quadratic.

\section{Ghost-gluon vertex.}

We repeat this exercise now for the structure of the ghost-gluon vertex and the
associated MOMh renormalization scheme. Though there are fewer amplitudes than
for the triple gluon vertex. For the $\MSbar$ scheme the two amplitudes are 
\begin{eqnarray}
\left. \Sigma^{\mbox{\footnotesize{ccg}}}_{(1)}(p,q) \right|_{\MSbars} &=& 
-~ 1 ~-~ \left[ 2.2324710 + 0.3280464 \alpha - 0.1464920 \alpha^2 \right] a 
\nonumber \\
&& -~ \left[ 49.2999213 + 16.6398011 \alpha + 2.3538026 \alpha^2
+ 0.0885284 \alpha^3 \right. \nonumber \\
&& \left. ~~~~+ 0.1098707 \alpha^4 ~-~ [ 4.0701546 + 0.4453256 \alpha
+ 0.1627713 \alpha^2 ] \Nf \right] a^2 \nonumber \\
&& +~ O(a^3) \nonumber \\
\left. \Sigma^{\mbox{\footnotesize{ccg}}}_{(2)}(p,q) \right|_{\MSbars} &=& 
\left[ 1.4824710 - 0.1640232 \alpha - 0.1464920 \alpha^2 \right] a 
\nonumber \\
&& +~ \left[ 35.1253580 + 2.2852097 \alpha - 0.3277110 \alpha^2
+ 0.0885284 \alpha^3 \right. \nonumber \\
&& \left. ~~~~+ 0.1098707 \alpha^4 ~-~ [ 3.2368212 + 0.4453256 \alpha
+ 0.1627713 \alpha^2 ] \Nf \right] a^2 \nonumber \\
&& +~ O(a^3) 
\end{eqnarray}
leading to the MOMh scheme expressions 
\begin{eqnarray}
\left. \Sigma^{\mbox{\footnotesize{ccg}}}_{(1)}(p,q) \right|_{\MOMhs} &=& 
-~ 1 ~+~ O(a^3) \nonumber \\
\left. \Sigma^{\mbox{\footnotesize{ccg}}}_{(2)}(p,q) \right|_{\MOMhs} &=& 
\left[ 1.4824710 - 0.1640232 \alpha - 0.1464920 \alpha^2 \right] a 
\nonumber \\
&& +~ \left[ 4.3185039 + 0.6852156 \alpha + 0.0493136 \alpha^2
- 0.0591276 \alpha^3 \right. \nonumber \\
&& \left. ~~~~- 0.0643817 \alpha^4 ~-~ [ 1.5896312 + 0.4453256 \alpha ] \Nf 
\right] a^2 \nonumber \\
&& +~ O(a^3) ~. 
\end{eqnarray}
As channel $1$ contained the poles in $\epsilon$ after MOMh renormalization 
then in that scheme there are no corrections at the symmetric subtraction 
point. Unlike the other two vertices we do not include a table comparing our
numerical expressions with those of \cite{22} as there is no direct 
comparison\footnote{We are grateful to Prof. Chetyrkin for comments on this 
point.}. Moreover, we have examined whether various combinations of our 
amplitudes can produce agreement, because of potentially different conventions
of defining the basis, but have failed to find any. For other quantities 
related to the MOMh scheme such as the coupling constant mapping and some 
coefficients of the three loop $SU(3)$ Landau gauge $\beta$-function we find 
similar but minor discrepancies with \cite{22} which suggest a consistent 
typographical error in the presentation of certain equations in \cite{22}. 
Further, related inconsistencies for the results for this vertex given in
\cite{22} have also been noted in \cite{28}. Whilst we will comment in more 
depth in section $7$ in the context of the results of the other two schemes we 
note that our benchmark check, \cite{24}, on the Landau gauge $SU(3)$ 
$\beta$-function for MOMh was slighty {\em better} than that of the MOMggg 
case.

Continuing with the presentation of our results, the $SU(3)$ numerical values 
for the respective MOMh conversion functions to $\MSbar$ are  
\begin{eqnarray}
C_g^{\MOMhs} (a,\alpha) &=& 1 ~-~ \left[ 9.2741377
+ 1.0780464 \alpha + 0.2285058 \alpha^2 - 0.5555556 \Nf \right] a \nonumber \\
&& -~ \left[ 191.9555891 + 18.6287007 \alpha + 1.8982822 \alpha^2
+ 0.7339954 \alpha^3 \right. \nonumber \\
&& \left. ~~~~-~ 0.0675921 \alpha^4 ~-~ [ 27.3210789 - 0.1535891 \alpha 
- 0.3808430 \alpha^2 ] \Nf \right. \nonumber \\
&& \left. ~~~~+~ 0.1543210 \Nf^2 \right] a^2 ~+~ O(a^3) \nonumber \\
C_A^{\MOMhs} (a,\alpha) &=& 1 ~+~ \left[ 8.0833333 
+ 1.5000000 \alpha + 0.7500000 \alpha^2 - 1.1111111 \Nf \right] a \nonumber \\
&& +~ \left[ 256.1034914 + 31.4182743 \alpha + 8.4375000 \alpha^2
+ 2.8125000 \alpha^3 \right] \nonumber \\
&& \left. ~~~~-~ [ 53.9129277 - 1.6666667 \alpha ] \Nf ~+~ 1.2345679 \Nf^2 
\right] a^2 ~+~ O(a^3) \nonumber \\
C_c^{\MOMhs} (a,\alpha) &=& 1 ~+~ 3.0000000 a \nonumber \\
&& +~ \left[ 80.9357699 + 3.0725670 \alpha + 1.3465290 \alpha^2 - 5.9375000 \Nf
\right] a^2 \nonumber \\
&& +~ O(a^3) \nonumber \\
C_\psi^{\MOMhs} (a,\alpha) &=& 1 ~-~ 1.3333333 \alpha a \nonumber \\
&& -~ \left[ 25.4642061 + 11.5753172 \alpha + 2.7222222 \alpha^2 
- 2.3333333 \Nf \right] a^2 \nonumber \\
&& +~ O(a^3) ~. 
\end{eqnarray}
The analytic form of the first of these leads to the coupling constant mapping 
\begin{eqnarray}
a_{\MOMhs} &=& a_{\MSbars} ~+~ \left[ \left[ 15 \psi^\prime(\third) - 10 \pi^2 
+ 615 \right] C_A ~-~ 240 T_F \Nf \right] \frac{a_{\MSbars}^2}{108} 
\nonumber \\ 
&& +~ \left[ \left[ 
450 (\psi^\prime(\third))^2
- 600 \pi^2 \psi^\prime(\third) 
- 458928 \psi^\prime(\third)
- 3213 \psi^{\prime\prime\prime}(\third) 
\right. \right. \nonumber \\
&& \left. \left. ~~~~~
-~ 3825792 s_2(\pisix)
+ 7651584 s_2(\pitwo)
+ 6376320 s_3(\pisix)
- 5101056 s_3(\pitwo)
\right. \right. \nonumber \\
&& \left. \left. ~~~~~
+~ 8768 \pi^4 
+ 305952 \pi^2 
+ 7776 \Sigma 
+ 153576 \zeta(3) 
+ 6521760
\right. \right. \nonumber \\
&& \left. \left. ~~~~~
-~ 26568 \frac{\ln^2(3) \pi}{\sqrt{3}} 
+ 318816 \frac{\ln(3) \pi}{\sqrt{3}}
+ 28536 \frac{\pi^3}{\sqrt{3}}
\right] C_A^2 ~+~ 460800 T_F^2 \Nf^2 
\right. \nonumber \\
&& ~~~~~+ \left. \left[ 
206784 \psi^\prime(\third)
+ 1492992 s_2(\pisix)
- 2985984 s_2(\pitwo)
- 2488320 s_3(\pisix)
\right. \right. \nonumber \\
&& \left. \left. ~~~~~~~~~
+~ 1990656 s_3(\pitwo)
- 137856 \pi^2 
- 995328 \zeta(3) 
- 4015296
\right. \right. \nonumber \\
&& \left. \left. ~~~~~~~~~
+~ 10368 \frac{\ln^2(3) \pi}{\sqrt{3}} 
- 124416 \frac{\ln(3) \pi}{\sqrt{3}}
- 11136 \frac{\pi^3}{\sqrt{3}}
\right] C_A T_F \Nf \right. \nonumber \\
&& \left. ~~~~~+ \left[ 
1492992 \zeta(3)
- 1710720 
\right] C_F T_F \Nf \right] \frac{a_{\MSbars}^3}{93312} ~+~ 
O\left( a_{\MSbars}^4 \right) 
\end{eqnarray}
or numerically, for an arbitrary linear covariant gauge, 
\begin{eqnarray}
a_{\MOMhs} &=& a_{\MSbars} ~+~ \left[ 18.5482754 + 2.1560928 \alpha_{\MSbars} 
+ 0.4570116 \alpha_{\MSbars}^2 ~-~ 1.1111111 \Nf \right] a_{\MSbars}^2 
\nonumber \\
&& +~ \left[ 641.9400674 + 97.2451047 \alpha_{\MSbars} 
+ 19.9982818 \alpha_{\MSbars}^2 + 2.9460299 \alpha_{\MSbars}^3 
\right. \nonumber \\
&& \left. ~~~~+ 0.0214606 \alpha_{\MSbars}^4 ~-~ \left[ 85.5559502 
+ 3.2863098 \alpha_{\MSbars} \right] \Nf \right. \nonumber \\
&& \left. ~~~~+ 1.2345679 \Nf^2 \right] a_{\MSbars}^3 ~+~ 
O\left( a_{\MSbars}^4 \right) ~.  
\end{eqnarray}
Using the renormalization group the three loop renormalization group functions  
emerge. Analytically in the Landau gauge we have 
\begin{eqnarray}
\beta^{\mbox{$\MOMhs$}}(a,0) &=& -~ \left[ \frac{11}{3} C_A - \frac{4}{3} 
T_F \Nf \right] a^2 ~-~ \left[ \frac{34}{3} C_A^2 - 4 C_F T_F \Nf 
- \frac{20}{3} C_A T_F \Nf \right] a^3 \nonumber \\
&& +~ \left[ \left[ 
4950 (\psi^\prime(\third))^2
- 6600 \pi^2 \psi^\prime(\third) 
+ 2370816 \psi^\prime(\third)
\right. \right. \nonumber \\
&& \left. \left. ~~~~~
+~ 11781 \psi^{\prime\prime\prime}(\third) 
+ 14027904 s_2(\pisix)
- 28055808 s_2(\pitwo)
- 23379840 s_3(\pisix)
\right. \right. \nonumber \\
&& \left. \left. ~~~~~
+~ 18703872 s_3(\pitwo)
- 29216 \pi^4 
- 1580544 \pi^2 
- 28512 \Sigma 
- 563112 \zeta(3) 
\right. \right. \nonumber \\
&& \left. \left. ~~~~~
-~ 11733336
+ 97416 \frac{\ln^2(3) \pi}{\sqrt{3}} 
- 1168992 \frac{\ln(3)\pi}{\sqrt{3}} 
- 104632 \frac{\pi^3}{\sqrt{3}}
\right] C_A^3 \right. \nonumber \\
&& \left. ~~~~~+~ \left[ 
2400 \pi^2 \psi^\prime(\third)
- 1800 (\psi^\prime(\third))^2
- 1864512 \psi^\prime(\third)
- 4284 \psi^{\prime\prime\prime}(\third) 
\right. \right. \nonumber \\
&& \left. \left. ~~~~~~~~~~
-~ 10575360 s_2(\pisix)
+ 21150720 s_2(\pitwo)
+ 17625600 s_3(\pisix)
\right. \right. \nonumber \\
&& \left. \left. ~~~~~~~~~~
-~ 14100480 s_3(\pitwo)
+ 10624 \pi^4 
+ 1243008 \pi^2 
+ 10368 \Sigma 
+ 3854304 \zeta(3) 
\right. \right. \nonumber \\
&& \left. \left. ~~~~~~~~~~
+~ 9722592
- 73440 \frac{\ln^2(3) \pi}{\sqrt{3}} 
+ 881280 \frac{\ln(3)\pi}{\sqrt{3}} 
+ 78880 \frac{\pi^3}{\sqrt{3}}
\right] C_A^2 T_F \Nf \right. \nonumber \\
&& \left. ~~~~~+~ \left[ 
352512 \psi^\prime(\third)
+ 1990656 s_2(\pisix)
- 3981312 s_2(\pitwo)
- 3317760 s_3(\pisix)
\right. \right. \nonumber \\
&& \left. \left. ~~~~~~~~~~
+ 2654208 s_3(\pitwo)
- 235008 \pi^2 
- 1327104 \zeta(3) 
- 1368576
\right. \right. \nonumber \\
&& \left. \left. ~~~~~~~~~~
+~ 13824 \frac{\ln^2(3) \pi}{\sqrt{3}} 
- 165888 \frac{\ln(3)\pi}{\sqrt{3}} 
- 14848 \frac{\pi^3}{\sqrt{3}}
\right] C_A T_F^2 \Nf^2 \right. \nonumber \\
&& \left. ~~~~~+~ \left[ 
34560 \pi^2
- 51840 \psi^\prime(\third)
- 5474304 \zeta(3)
+ 6272640
\right] C_A C_F T_F \Nf \right. \nonumber \\
&& \left. ~~~~~+~ \left[ 
1990656 \zeta(3)
- 1907712
\right] C_F T_F^2 \Nf^2 - 186624 C_F^2 T_F \Nf \right] \frac{a^4}{93312}
\nonumber \\
&& +~ O(a^5)
\end{eqnarray}
\begin{eqnarray}
\gamma_A^{\mbox{$\MOMhs$}}(a,0) &=& -~ \left[ 13 C_A - 8 T_F \Nf \right] 
\frac{a}{6} \nonumber \\
&& + \left[ \left[ 195 \psi^\prime(\third) - 130 \pi^2 - 3186 \right] C_A^2
+ 2592 C_F T_F \Nf \right. \nonumber \\
&& \left. ~~~+ \left[ 80 \pi^2 - 120 \psi^\prime(\third) + 2808 \right] 
C_A T_F \Nf \right] \frac{a^2}{648} \nonumber \\ 
&& +~ \left[ \left[ 
54600 \pi^2 \psi^\prime(\third) 
- 40950 (\psi^\prime(\third))^2
- 7120224 \psi^\prime(\third)
- 41769 \psi^{\prime\prime\prime}(\third) 
\right. \right. \nonumber \\
&& \left. \left. ~~~~~
-~ 49735296 s_2(\pisix)
+ 99470592 s_2(\pitwo)
+ 82892160 s_3(\pisix)
- 66313728 s_3(\pitwo)
\right. \right. \nonumber \\
&& \left. \left. ~~~~~
+~ 93184 \pi^4 
+ 4746816 \pi^2 
+ 101088 \Sigma 
+ 14628600 \zeta(3) 
- 37070136
\right. \right. \nonumber \\
&& \left. \left. ~~~~~
-~ 345384 \frac{\ln^2(3) \pi}{\sqrt{3}} 
+ 1414608 \frac{\ln(3) \pi}{\sqrt{3}}
+ 370968 \frac{\pi^3}{\sqrt{3}}
\right] C_A^3 \right. \nonumber \\
&& ~~~~~+ \left. \left[ 
25200 (\psi^\prime(\third))^2
- 33600 \pi^2 \psi^\prime(\third) 
+ 7615296 \psi^\prime(\third)
+ 25704 \psi^{\prime\prime\prime}(\third) 
\right. \right. \nonumber \\
&& \left. \left. ~~~~~~~~~
+~ 50015232 s_2(\pisix)
- 100030464 s_2(\pitwo)
- 83358720 s_3(\pisix)
\right. \right. \nonumber \\
&& \left. \left. ~~~~~~~~~
+~ 66686976 s_3(\pitwo)
- 57344 \pi^4 
- 5076864 \pi^2 
- 62208 \Sigma 
\right. \right. \nonumber \\
&& \left. \left. ~~~~~~~~~
+~ 4121280 \zeta(3) 
+ 35848360
+ 347328 \frac{\ln^2(3) \pi}{\sqrt{3}} 
\right. \right. \nonumber \\
&& \left. \left. ~~~~~~~~~
-~ 4167936 \frac{\ln(3) \pi}{\sqrt{3}}
- 373056 \frac{\pi^3}{\sqrt{3}}
\right] C_A^2 T_F \Nf ~-~ 1119744 C_F^2 T_F \Nf \right. \nonumber \\
&& ~~~~~+ \left. \left[ 
23887872 s_2(\pitwo)
- 2115072 \psi^\prime(\third)
- 11943936 s_2(\pisix)
+ 19906560 s_3(\pisix)
\right. \right. \nonumber \\
&& \left. \left. ~~~~~~~~~
-~ 15925248 s_3(\pitwo)
+ 1410048 \pi^2 
- 3981312 \zeta(3) 
- 5225472
\right. \right. \nonumber \\
&& \left. \left. ~~~~~~~~~
-~ 82944 \frac{\ln^2(3) \pi}{\sqrt{3}} 
+ 995328 \frac{\ln(3) \pi}{\sqrt{3}}
+ 89088 \frac{\pi^3}{\sqrt{3}}
\right] C_A T_F^2 \Nf^2 \right. \nonumber \\
&& ~~~~~+ \left. \left[ 
414720 \pi^2 
- 622080 \psi^\prime(\third)
- 32845824 \zeta(3) 
+ 33716736
\right] C_A C_F T_F \Nf \right. \nonumber \\
&& ~~~~~+ \left. \left[ 
11943936 \zeta(3) 
- 11446272
\right] C_F T_F^2 \Nf^2 \right] \frac{a^3}{559872} ~+~ O(a^4)
\end{eqnarray}
\begin{eqnarray}
\gamma_c^{\mbox{$\MOMhs$}}(a,0) &=& -~ \frac{3}{4} C_A a ~+~ \left[ 
\left[ 15 \psi^\prime(\third) - 10 \pi^2 - 198 \right] C_A ~+~ 
72 T_F \Nf \right] \frac{C_A a^2}{144} \nonumber \\ 
&& +~ \left[ \left[ 
4200 \pi^2 \psi^\prime(\third) 
- 3150 (\psi^\prime(\third))^2
- 559008 \psi^\prime(\third)
- 3213 \psi^{\prime\prime\prime}(\third) 
\right. \right. \nonumber \\
&& \left. \left. ~~~~~
-~ 3825792 s_2(\pisix)
+ 7651584 s_2(\pitwo)
+ 6376320 s_3(\pisix)
- 5101056 s_3(\pitwo)
\right. \right. \nonumber \\
&& \left. \left. ~~~~~
+~ 7168 \pi^4 
+ 372672 \pi^2 
+ 7776 \Sigma 
+ 973944 \zeta(3) 
- 2855736
\right. \right. \nonumber \\
&& \left. \left. ~~~~~
-~ 26568 \frac{\ln^2(3) \pi}{\sqrt{3}} 
+ 318816 \frac{\ln(3) \pi}{\sqrt{3}}
+ 28536 \frac{\pi^3}{\sqrt{3}}
\right] C_A^2 \right. \nonumber \\
&& ~~~~~+ \left. \left[ 
247104 \psi^\prime(\third)
+ 1492992 s_2(\pisix)
- 2985984 s_2(\pitwo)
- 2488320 s_3(\pisix)
\right. \right. \nonumber \\
&& \left. \left. ~~~~~~~~~
+~ 1990656 s_3(\pitwo)
- 164736 \pi^2 
- 186624 \zeta(3) 
+ 2260224
\right. \right. \nonumber \\
&& \left. \left. ~~~~~~~~~
+~ 10368 \frac{\ln^2(3) \pi}{\sqrt{3}} 
- 124416 \frac{\ln(3) \pi}{\sqrt{3}}
- 11136 \frac{\pi^3}{\sqrt{3}}
\right] C_A T_F \Nf \right. \nonumber \\
&& \left. ~~~~~
-~ 414720 T_F^2 \Nf^2 ~+~ 186624 C_F T_F \Nf \right] \frac{C_A a^3}{124416} ~+~
O(a^4) 
\end{eqnarray}
and 
\begin{eqnarray}
\gamma_\psi^{\mbox{$\MOMhs$}}(a,0) &=& \left[ 25 C_A - 6 C_F 
- 8 T_F \Nf \right] \frac{C_F a^2}{4} \nonumber \\
&& +~ \left[ \left[ 500 \pi^2 - 750 \psi^\prime(\third)
- 13230 \zeta(3) + 29187 \right] C_A^2 \right. \nonumber \\
&& \left. ~~~~+ \left[ 180 \psi^\prime(\third) - 120 \pi^2 
+ 5184 \zeta(3) - 10044 \right] C_A C_F \right. \nonumber \\
&& \left. ~~~~+ \left[ 240 \psi^\prime(\third) - 160 \pi^2 
+ 3456 \zeta(3) - 14832 \right] C_A T_F \Nf \right. \nonumber \\
&& \left. ~~~~-~ 864 C_F T_F \Nf ~+~ 648 C_F^2 ~+~ 1152 T_F^2 \Nf^2 
\right] \frac{C_F a^2}{432} ~+~ O(a^4) ~. 
\end{eqnarray}
For comparison with $\MSbar$ and the other two MOM schemes we note the full
arbitrary linear gauge values are 
\begin{eqnarray}
\beta^{\mbox{$\MOMhs$}}(a,\alpha) &=& -~ [ 11.0000000 - 0.6666667 \Nf ] a^2
\nonumber \\
&& -~ \left[ 102.0000000 - 14.0146029 \alpha - 2.7070116 \alpha^2
+ 1.3710348 \alpha^3 \right. \nonumber \\
&& \left. ~~~~- \left[ 12.6666667 - 1.4373952 \alpha - 0.6093488 \alpha^2
\right] \Nf \right] a^3 \nonumber \\
&& -~ \left[ 2813.4929483 - 138.6933593 \alpha - 35.8596355 \alpha^2
+ 4.4867569 \alpha^3 \right. \nonumber \\
&& \left. ~~~~+ 9.6466346 \alpha^4 + 1.3338552 \alpha^5 \right. \nonumber \\
&& \left. ~~~~- \left[ 617.6471542 - 40.9610609 \alpha - 14.8518881 \alpha^2
- 2.2357888 \alpha^3 \right. \right. \nonumber \\
&& \left. \left. ~~~~~~~~- 0.2607456 \alpha^4 \right] \Nf ~+~ 21.5028181 
\Nf^2 \right] a^4 ~+~ O(a^5) \nonumber \\
\gamma_A^{\mbox{$\MOMhs$}}(a,\alpha) &=& [ 0.6666667 \Nf - 6.5000000
+ 1.5000000 \alpha ] a \nonumber \\
&& -~ \left[ 34.7278768 - 3.9421899 \alpha - 3.4864362 \alpha^2
+ 1.8105174 \alpha^3 \right. \nonumber \\
&& \left. ~~~~- \left[ 8.1900386 - 1.4373952 \alpha - 0.8046744 \alpha^2
\right] \Nf \right] a^2 \nonumber \\
&& -~ \left[ 1195.1013833 - 28.3678253 \alpha - 8.6742911 \alpha^2
- 5.5422477 \alpha^3 \right. \nonumber \\
&& \left. ~~~~+ 10.3727415 \alpha^4 + 1.0931120 \alpha^5 \right. \nonumber \\
&& \left. ~~~~- \left[ 323.9161114 - 37.7892337 \alpha - 11.8481326 \alpha^2
- 1.7973717 \alpha^3 \right. \right. \nonumber \\
&& \left. \left. ~~~~~~~~- 0.1108276 \alpha^4 \right] \Nf ~+~ \left[ 
12.2638016 - 0.5937675 \alpha \right] \Nf^2 \right] a^3 ~+~ O(a^4) \nonumber \\
\gamma_c^{\mbox{$\MOMhs$}}(a,\alpha) &=& [ 0.7500000 \alpha - 2.2500000 ] a 
\nonumber \\
&& -~ \left[ 9.0788804 + 3.5599978 \alpha - 0.5362065 \alpha^2
- 0.2197413 \alpha^3 \right. \nonumber \\
&& \left. ~~~~- 0.7500000 \Nf \right] a^2 \nonumber \\
&& -~ \left[ 462.6953814 - 89.8527064 \alpha - 1.2245066 \alpha^2
+ 2.9101869 \alpha^3 \right. \nonumber \\
&& \left. ~~~~- 0.3430542 \alpha^4 - 0.1126679 \alpha^5 \right. \nonumber \\
&& \left. ~~~~- \left[ 76.4163746 - 1.4615295 \alpha - 0.0175290 \alpha^2
\right] \Nf ~+~ 2.5000000 \Nf^2 \right] a^3 \nonumber \\
&& +~ O(a^4) \nonumber \\
\gamma_\psi^{\mbox{$\MOMhs$}}(a,\alpha) &=& 1.3333333 \alpha a \nonumber \\
&& +~ \left[ 22.3333333 + 0.0467440 \alpha + 2.1252097 \alpha^2
+ 0.3906512 \alpha^3 \right. \nonumber \\
&& \left. ~~~~- 1.3333333 \Nf \right] a^2 \nonumber \\
&& +~ \left[ 260.0472082 - 162.9606897 \alpha - 38.3957984 \alpha^2
+ 2.9734643 \alpha^3 \right. \nonumber \\
&& \left. ~~~~+ 1.2107696 \alpha^4 + 0.2002985 \alpha^5 \right. \nonumber \\
&& \left. ~~~~- \left[ 47.3050219 - 12.9383858 \alpha - 2.4062325 \alpha^2
\right] \Nf ~+~ 0.8888889 \Nf^2 \right] a^3 \nonumber \\
&& +~ O(a^4) ~.  
\end{eqnarray}
The $\beta$-function has the same $\Nf$ polynomial structure as the $\MSbar$
scheme.

\section{Quark-gluon vertex.}

Finally, we complete our task with the quark-gluon vertex and the MOMq scheme
expressions. The $\MSbar$ amplitudes are 
\begin{eqnarray}
\left. \Sigma^{\mbox{\footnotesize{qqv}}}_{(1)}(p,q) \right|_{\MSbars} &=& 
1 ~+~ \left[ 4.3162206 - 0.5887601 \alpha - 0.4570116 \alpha^2 \right] a 
\nonumber \\
&& +~ \left[ 89.2876778 - 2.5488660 \alpha + 0.7959457 \alpha^2
+ 0.2344278 \alpha^3 \right. \nonumber \\
&& \left. ~~~~+ 0.3427587 \alpha^4 ~-~ [ 12.1366772 + 0.9766280 \alpha
+ 0.5077907 \alpha^2 ] \Nf \right] a^2 \nonumber \\
&& +~ O(a^3) \nonumber \\
\left. \Sigma^{\mbox{\footnotesize{qqv}}}_{(2)}(p,q) \right|_{\MSbars} &=& 
\left. \Sigma^{\mbox{\footnotesize{qqv}}}_{(5)}(p,q) \right|_{\MSbars} 
\nonumber \\
&=& \left[ 2.5980335 - 2.3056953 \alpha - 0.4140232 \alpha^2 \right] a 
\nonumber \\
&& +~ \left[ 26.4812470 - 21.7488508 \alpha - 5.3984938 \alpha^2
+ 0.4547874 \alpha^3 \right. \nonumber \\
&& \left. ~~~~+ 0.3105174 \alpha^4 ~-~ [ 6.2718940 + 1.0339459 \alpha
+ 0.4600258 \alpha^2 ] \Nf \right] a^2 \nonumber \\
&& +~ O(a^3) \nonumber \\
\left. \Sigma^{\mbox{\footnotesize{qqv}}}_{(3)}(p,q) \right|_{\MSbars} &=& 
\left. \Sigma^{\mbox{\footnotesize{qqv}}}_{(4)}(p,q) \right|_{\MSbars} 
\nonumber \\
&=& \left[ 2.0502686 - 2.5226306 \alpha - 0.5000000 \alpha^2 \right] a 
\nonumber \\
&& +~ \left[ 12.7352941 - 25.2299763 \alpha - 6.6819786 \alpha^2
+ 0.0320681 \alpha^3 \right. \nonumber \\
&& \left. ~~~~+ 0.3750000 \alpha^4 ~-~ [ 4.8715927 + 0.9193101 \alpha
+ 0.5555555 \alpha^2 ] \Nf \right] a^2 \nonumber \\
&& +~ O(a^3) \nonumber \\
\left. \Sigma^{\mbox{\footnotesize{qqv}}}_{(6)}(p,q) \right|_{\MSbars} &=& -~ 
\left[ 4.3622718 + 2.3439072 \alpha + 0.5859768 \alpha^2 \right] a 
\nonumber \\
&& -~ \left[ 131.9911153 + 45.4675027 \alpha + 4.8573516 \alpha^2
+ 1.2207850 \alpha^3 \right. \nonumber \\
&& \left. ~~~~- 0.4394826 \alpha^4 ~-~ [ 10.9228497 + 1.9532560 \alpha
- 0.6510853 \alpha^2 ] \Nf \right] a^2 \nonumber \\
&& +~ O(a^3) 
\end{eqnarray}
where the symmetry of the exchange of the two external quark legs is manifest.
This emerged from the computation and was not imposed. In Table $2$ we have 
given the comparison table analogous to Table $1$ for this vertex. Again the
numerical estimates and the results we have produced are virtually identical. 

{\begin{table}[ht]
\begin{center}
\begin{tabular}{|c||r|r|r|r|r|}
\hline
$\left. \Sigma^{\mbox{\footnotesize{qqv}}}_{(1),2}(p,q) \right|_{\MSbars}$ 
& $C_F^2$ & $C_F C_A$ & $C_A^2$ & $C_A T_F \Nf$ & $C_F T_F \Nf$ \\
\hline
Ref \cite{22} & $0.206(4)$ & $-$ $0.20(4)$ & $0.679(1)$ & $-$ $0.4968(4)$ & 
$-$ $0.0211(4)$ \\ 
This paper & $0.2048069$ & $-$ $0.2158263$ & $0.6755204$ & $-$ $0.4958508$ &
$-$ $0.0221492$ \\ 
\hline
\end{tabular}
\end{center}
\begin{center}
{Table $2$. Comparison of channel $1$ two loop Landau gauge amplitude with
\cite{22} by colour factor.}  
\end{center}
\end{table}}

The corresponding MOMq scheme expressions are
\begin{eqnarray}
\left. \Sigma^{\mbox{\footnotesize{qqv}}}_{(1)}(p,q) \right|_{\MOMqs} &=& 
1 ~+~ O(a^3) \nonumber \\
\left. \Sigma^{\mbox{\footnotesize{qqv}}}_{(2)}(p,q) \right|_{\MOMqs} &=& 
\left. \Sigma^{\mbox{\footnotesize{qqv}}}_{(5)}(p,q) \right|_{\MOMqs} 
\nonumber \\
&=& \left[ 2.5980335 - 2.3056953 \alpha - 0.4140232 \alpha^2 \right] a 
\nonumber \\
&& -~ \left[ 28.1605810 - 15.7267125 \alpha + 11.9916906 \alpha^2
+ 5.1627786 \alpha^3 \right. \nonumber \\
&& \left. ~~~~+ 0.5676402 \alpha^4 ~+~ [ 3.3851901 + 1.0339459 \alpha) \Nf 
\right] a^2 ~+~ O(a^3) \nonumber \\
\left. \Sigma^{\mbox{\footnotesize{qqv}}}_{(3)}(p,q) \right|_{\MOMqs} &=& 
\left. \Sigma^{\mbox{\footnotesize{qqv}}}_{(4)}(p,q) \right|_{\MOMqs} 
\nonumber \\
&=& \left[ 2.0502686 - 2.5226306 \alpha - 0.5000000 \alpha^2 \right] a 
\nonumber \\
&& -~ \left[ 30.3859453 - 13.4480431 \alpha + 14.1587135 \alpha^2
+ 6.3930196 \alpha^3 \right. \nonumber \\
&& \left. ~~~~+ 0.6855174 \alpha^4 ~+~ [ 2.5935165 + 0.9193101 \alpha) \Nf 
\right] a^2 ~+~ O(a^3) \nonumber \\
\left. \Sigma^{\mbox{\footnotesize{qqv}}}_{(6)}(p,q) \right|_{\MOMqs} &=& -~ 
\left[ 4.3622718 + 2.3439072 \alpha + 0.5859768 \alpha^2 \right] a 
\nonumber \\
&& -~ \left[ 40.2438356 + 27.9113517 \alpha + 15.1059206 \alpha^2
+ 7.9109326 \alpha^3 \right. \nonumber \\
&& \left. ~~~~+ 0.8033946 \alpha^4 ~-~ [ 6.0758810 + 1.9532560 \alpha) \Nf 
\right] a^2 \nonumber \\
&& +~ O(a^3) 
\end{eqnarray}
where clearly channel $1$ correctly corresponds to the MOMq scheme definition 
as it is the only channel to contain the divergences in $\epsilon$. The quark 
external leg interchange also correctly emerges as a minor check on our 
computations corresponding to swapping $p$ and $q$. From the renormalization 
constants in each scheme the numerical conversion functions from MOMq to 
$\MSbar$ for $SU(3)$ are
\begin{eqnarray}
C_g^{\MOMqs} (a,\alpha) &=& 1 ~-~ \left[ 8.3578873 
- 1.1720934 \alpha - 0.0820116 \alpha^2 - 0.5555556 Nf \right] a \nonumber \\
&& -~ \left[ 131.2981279 + 7.8598968 \alpha - 0.3923795 \alpha^2
+ 0.7219823 \alpha^3 \right. \nonumber \\
&& \left. ~~~~+~ 0.0943409 \alpha^4 ~-~ [ 27.6257962 + 1.6277910 \alpha
+ 0.1366860 \alpha^2 ] \Nf \right. \nonumber \\
&& \left. ~~~~+~ 0.1543230 \Nf^2 \right] a^2 ~+~ O(a^3) \nonumber \\
C_A^{\MOMqs} (a,\alpha) &=& 1 ~+~ \left[ 8.0833333 
+ 1.5000000 \alpha + 0.7500000 \alpha^2 - 1.1111111 \Nf \right] a \nonumber \\
&& +~ \left[ 256.1034914 + 31.4182743 \alpha + 8.4375000 \alpha^2
+ 2.8125000 \alpha^3 \right] \nonumber \\
&& \left. ~~~~-~ [ 53.9129277 - 1.6666667 \alpha ] \Nf ~+~ 1.2345679 \Nf^2 
\right] a^2 ~+~ O(a^3) \nonumber \\
C_c^{\MOMqs} (a,\alpha) &=& 1 ~+~ 3.0000000 a \nonumber \\
&& +~ \left[ 80.9357699 + 3.0725670 \alpha + 1.3465290 \alpha^2 - 5.9375000 \Nf
\right] a^2 \nonumber \\
&& +~ O(a^3) \nonumber \\
C_\psi^{\MOMqs} (a,\alpha) &=& 1 ~-~ 1.3333333 \alpha a \nonumber \\
&& -~ \left[ 25.4642061 + 11.5753172 \alpha + 2.7222222 \alpha^2 
- 2.3333333 \Nf \right] a^2 \nonumber \\
&& +~ O(a^3) ~. 
\end{eqnarray}
The analytic form of the first produces the coupling constant mapping which is
\begin{eqnarray}
a_{\MOMqs} &=& a_{\MSbars} \nonumber \\
&& +~ \left[ \left[ 52 \pi^2 - 78 \psi^\prime(\third) + 993 \right] C_A ~-~
240 T_F \Nf ~+~ \left[ 48 \psi^\prime(\third) - 32 \pi^2 - 432 \right] C_F
\right] \frac{a_{\MSbars}^2}{108} \nonumber \\ 
&& +~ \left[ \left[ 
9768 \pi^2 \psi^\prime(\third) 
- 7326 (\psi^\prime(\third))^2
- 350910 \psi^\prime(\third)
+ 2133 \psi^{\prime\prime\prime}(\third) 
\right. \right. \nonumber \\
&& \left. \left. ~~~~~
+~ 664848 s_2(\pisix)
- 1329696 s_2(\pitwo)
- 1108080 s_3(\pisix)
+ 886464 s_3(\pitwo)
\right. \right. \nonumber \\
&& \left. \left. ~~~~~
-~ 8944 \pi^4 
+ 233940 \pi^2 
- 54432 \Sigma 
- 164754 \zeta(3) 
+ 3278628
\right. \right. \nonumber \\
&& \left. \left. ~~~~~
+~ 4617 \frac{\ln^2(3) \pi}{\sqrt{3}} 
- 55404 \frac{\ln(3) \pi}{\sqrt{3}}
- 4959 \frac{\pi^3}{\sqrt{3}}
\right] C_A^2 \right. \nonumber \\
&& ~~~~~+ \left. \left[ 
27360 (\psi^\prime(\third))^2
- 36480 \pi^2 \psi^\prime(\third) 
- 174816 \psi^\prime(\third)
+ 1296 \psi^{\prime\prime\prime}(\third) 
\right. \right. \nonumber \\
&& \left. \left. ~~~~~~~~~
-~ 1959552 s_2(\pisix)
+ 3919104 s_2(\pitwo)
+~ 3265920 s_3(\pisix)
- 2612736 s_3(\pitwo)
\right. \right. \nonumber \\
&& \left. \left. ~~~~~~~~~
+~ 8704 \pi^4 
+ 116544 \pi^2 
+ 46656 \zeta(3) 
- 1960848
\right. \right. \nonumber \\
&& \left. \left. ~~~~~~~~~
-~ 13608 \frac{\ln^2(3) \pi}{\sqrt{3}} 
+ 163296 \frac{\ln(3) \pi}{\sqrt{3}}
+ 14616 \frac{\pi^3}{\sqrt{3}}
\right] C_A C_F \right. \nonumber \\
&& ~~~~~+ \left. \left[ 
26112 \pi^2 \psi^\prime(\third) 
- 19584 (\psi^\prime(\third))^2
+ 295488 \psi^\prime(\third)
- 7776 \psi^{\prime\prime\prime}(\third) 
\right. \right. \nonumber \\
&& \left. \left. ~~~~~~~~~
-~ 373248 s_2(\pisix)
+ 746496 s_2(\pitwo)
+ 622080 s_3(\pisix)
- 497664 s_3(\pitwo)
\right. \right. \nonumber \\
&& \left. \left. ~~~~~~~~~
+~ 12032 \pi^4 
-~ 196992 \pi^2 
+ 31104 \Sigma
+ 435456 \zeta(3) 
+ 256608
\right. \right. \nonumber \\
&& \left. \left. ~~~~~~~~~
-~ 2592 \frac{\ln^2(3) \pi}{\sqrt{3}} 
+ 31104 \frac{\ln(3) \pi}{\sqrt{3}}
+ 2784 \frac{\pi^3}{\sqrt{3}}
\right] C_F^2 \right. \nonumber \\
&& ~~~~~+ \left. \left[ 
197568 \psi^\prime(\third)
- 864 \psi^{\prime\prime\prime}(\third) 
+~ 186624 s_2(\pisix)
- 373248 s_2(\pitwo)
- 311040 s_3(\pisix)
\right. \right. \nonumber \\
&& \left. \left. ~~~~~~~~~
+~ 248832 s_3(\pitwo)
+ 2304 \pi^4 
- 131712 \pi^2 
- 217728 \zeta(3) 
- 1549440
\right. \right. \nonumber \\
&& \left. \left. ~~~~~~~~~
+~ 1296 \frac{\ln^2(3) \pi}{\sqrt{3}} 
- 15552 \frac{\ln(3) \pi}{\sqrt{3}}
- 1392 \frac{\pi^3}{\sqrt{3}}
\right] C_A T_F \Nf ~+~
115200 T_F^2 \Nf^2 \right. \nonumber \\
&& ~~~~~+ \left. \left[ 
44544 \pi^2 
- 66816 \psi^\prime(\third)
+ 373248 \zeta(3)
+ 80352 
\right] C_F T_F \Nf \right] \frac{a_{\MSbars}^3}{23328} \nonumber \\
&& +~ O\left( a_{\MSbars}^4 \right) 
\end{eqnarray}
in the Landau gauge. Numerically we have 
\begin{eqnarray}
a_{\MOMqs} &=& a_{\MSbars} ~+~ \left[ 16.7157746 - 2.3441868 \alpha_{\MSbars} 
- 0.1640232 \alpha_{\MSbars}^2 ~-~ 1.1111111 \Nf \right] a_{\MSbars}^2 
\nonumber \\
&& +~ \left[ 472.1590953 - 43.0575532 \alpha_{\MSbars} 
- 0.7760123 \alpha_{\MSbars}^2 + 2.0207162 \alpha_{\MSbars}^3 
\right. \nonumber \\
&& \left. ~~~~+ 0.2088596 \alpha_{\MSbars}^4 ~-~ \left[ 83.1112168 
- 0.6513959 \alpha_{\MSbars} \right] \Nf \right. \nonumber \\
&& \left. ~~~~+ 1.2345679 \Nf^2 \right] a_{\MSbars}^3 ~+~ O\left( a_{\MSbars}^4 
\right) ~. 
\end{eqnarray}
Thus, similar to the previous sections the explicit three loop Landau gauge 
renormalization group functions are  
\begin{eqnarray}
\beta^{\mbox{$\MOMqs$}}(a,0) &=& -~ \left[ \frac{11}{3} C_A - \frac{4}{3} 
T_F \Nf \right] a^2 ~-~ \left[ \frac{34}{3} C_A^2 - 4 C_F T_F \Nf 
- \frac{20}{3} C_A T_F \Nf \right] a^3 \nonumber \\
&& +~ \left[ \left[ 
71478 (\psi^\prime(\third))^2
- 95304 \pi^2 \psi^\prime(\third) 
- 40266 \psi^\prime(\third)
- 7821 \psi^{\prime\prime\prime}(\third) 
\right. \right. \nonumber \\
&& \left. \left. ~~~~~
-~ 2437776 s_2(\pisix)
+ 4875552 s_2(\pitwo)
+ 4062960 s_3(\pisix)
- 3250368 s_3(\pitwo)
\right. \right. \nonumber \\
&& \left. \left. ~~~~~
+~ 52624 \pi^4 
+ 26844 \pi^2 
+ 199584 \Sigma 
+ 604098 \zeta(3) 
- 3593970
\right. \right. \nonumber \\
&& \left. \left. ~~~~~
-~ 16929 \frac{\ln^2(3) \pi}{\sqrt{3}}  
+ 203148 \frac{\ln(3) \pi}{\sqrt{3}}
+ 18183 \frac{\pi^3}{\sqrt{3}} 
\right] C_A^3 \right. \nonumber \\
&& \left. ~~~~~+ \left[ 
206976 \pi^2 \psi^\prime(\third) 
- 155232 (\psi^\prime(\third))^2
+ 1951776 \psi^\prime(\third)
- 4752 \psi^{\prime\prime\prime}(\third) 
\right. \right. \nonumber \\
&& \left. \left. ~~~~~~~~~
+~ 7185024 s_2(\pisix)
- 14370048 s_2(\pitwo)
- 11975040 s_3(\pisix)
+ 9580032 s_3(\pitwo)
\right. \right. \nonumber \\
&& \left. \left. ~~~~~~~~~
-~ 56320 \pi^4 
- 1301184 \pi^2 
- 171072 \zeta(3) 
- 159408
\right. \right. \nonumber \\
&& \left. \left. ~~~~~~~~~
+~ 49896 \frac{\ln^2(3) \pi}{\sqrt{3}}  
- 598752 \frac{\ln(3) \pi}{\sqrt{3}}
- 53592 \frac{\pi^3}{\sqrt{3}} 
\right] C_A^2 C_F \right. \nonumber \\
&& \left. ~~~~~+ \left[ 
34656 \pi^2 \psi^\prime(\third) 
- 25992 (\psi^\prime(\third))^2
- 392328 \psi^\prime(\third)
+ 6012 \psi^{\prime\prime\prime}(\third) 
\right. \right. \nonumber \\
&& \left. \left. ~~~~~~~~~
+~ 202176 s_2(\pisix)
- 404352 s_2(\pitwo)
- 336960 s_3(\pisix)
+ 269568 s_3(\pitwo)
\right. \right. \nonumber \\
&& \left. \left. ~~~~~~~~~
-~ 27584 \pi^4 
+ 261552 \pi^2 
- 72576 \Sigma
+ 578664 \zeta(3) 
+ 3133080
\right. \right. \nonumber \\
&& \left. \left. ~~~~~~~~~
+~ 1404 \frac{\ln^2(3) \pi}{\sqrt{3}}  
- 16848 \frac{\ln(3) \pi}{\sqrt{3}}
- 1508 \frac{\pi^3}{\sqrt{3}} 
\right] C_A^2 T_F \Nf \right. \nonumber \\
&& \left. ~~~~~+ \left[ 
88704 (\psi^\prime(\third))^2
- 118272 \pi^2 \psi^\prime(\third) 
- 1387584 \psi^\prime(\third)
+ 28512 \psi^{\prime\prime\prime}(\third) 
\right. \right. \nonumber \\
&& \left. \left. ~~~~~~~~~
+~ 1368576 s_2(\pisix)
- 2737152 s_2(\pitwo)
- 2280960 s_3(\pisix)
+ 1824768 s_3(\pitwo)
\right. \right. \nonumber \\
&& \left. \left. ~~~~~~~~~
-~ 36608 \pi^4 
+ 925056 \pi^2 
- 114048 \Sigma
- 1596672 \zeta(3) 
+ 427680
\right. \right. \nonumber \\
&& \left. \left. ~~~~~~~~~
+~ 9504 \frac{\ln^2(3) \pi}{\sqrt{3}}  
- 114048 \frac{\ln(3) \pi}{\sqrt{3}}
- 10208 \frac{\pi^3}{\sqrt{3}} 
\right] C_A C_F^2 \right. \nonumber \\
&& \left. ~~~~~+ \left[ 
56448 (\psi^\prime(\third))^2
- 75264 \pi^2 \psi^\prime(\third) 
- 592704 \psi^\prime(\third)
+ 1728 \psi^{\prime\prime\prime}(\third) 
\right. \right. \nonumber \\
&& \left. \left. ~~~~~~~~~
-~ 2612736 s_2(\pisix)
+ 5225472 s_2(\pitwo)
+ 4354560 s_3(\pisix)
- 3483648 s_3(\pitwo)
\right. \right. \nonumber \\
&& \left. \left. ~~~~~~~~~
+~ 20480 \pi^4 
+ 395136 \pi^2 
- 1306368 \zeta(3) 
+ 1194912
\right. \right. \nonumber \\
&& \left. \left. ~~~~~~~~~
-~ 18144 \frac{\ln^2(3) \pi}{\sqrt{3}}  
+ 217728 \frac{\ln(3) \pi}{\sqrt{3}}
+ 19488 \frac{\pi^3}{\sqrt{3}} 
\right] C_A C_F T_F \Nf \right. \nonumber \\
&& \left. ~~~~~+ \left[ 
163584 \psi^\prime(\third)
- 1152 \psi^{\prime\prime\prime}(\third) 
+ 248832 s_2(\pisix)
- 497664 s_2(\pitwo)
\right. \right. \nonumber \\
&& \left. \left. ~~~~~~~~~
-~ 414720 s_3(\pisix)
+ 331776 s_3(\pitwo)
+ 3072 \pi^4 
- 109056 \pi^2 
\right. \right. \nonumber \\
&& \left. \left. ~~~~~~~~~
-~ 290304 \zeta(3) 
- 585792
+ 1728 \frac{\ln^2(3) \pi}{\sqrt{3}}  
\right. \right. \nonumber \\
&& \left. \left. ~~~~~~~~~
-~ 20736 \frac{\ln(3) \pi}{\sqrt{3}}
- 1856 \frac{\pi^3}{\sqrt{3}} 
\right] C_A T_F^2 \Nf^2 \right. \nonumber \\
&& \left. ~~~~~+ \left[ 
43008 \pi^2 \psi^\prime(\third)
- 32256 (\psi^\prime(\third))^2
+ 463104 \psi^\prime(\third)
- 10368 \psi^{\prime\prime\prime}(\third) 
\right. \right. \nonumber \\
&& \left. \left. ~~~~~~~~~
-~ 497664 s_2(\pisix)
+ 995328 s_2(\pitwo)
+ 829440 s_3(\pisix)
- 663552 s_3(\pitwo)
\right. \right. \nonumber \\
&& \left. \left. ~~~~~~~~~
+~ 13312 \pi^4 
- 308736 \pi^2 
+ 41472 \Sigma
+ 580608 \zeta(3) 
+ 171072
\right. \right. \nonumber \\
&& \left. \left. ~~~~~~~~~
-~ 3456 \frac{\ln^2(3) \pi}{\sqrt{3}}  
+ 41472 \frac{\ln(3) \pi}{\sqrt{3}}
+ 3712 \frac{\pi^3}{\sqrt{3}} 
\right] C_F^2 T_F \Nf \right. \nonumber \\
&& \left. ~~~~~+ \left[ 
18432 \pi^2
- 27648 \psi^\prime(\third)
+ 497664 \zeta(3)
- 352512
\right] C_F T_F^2  \Nf^2 \Nf \right] \frac{a^4}{23328} \nonumber \\
&& +~ O(a^5)
\end{eqnarray}
\begin{eqnarray}
\gamma_A^{\mbox{$\MOMqs$}}(a,0) &=& -~ \left[ 13 C_A - 8 T_F \Nf \right] 
\frac{a}{6} \nonumber \\
&& + \left[ \left[ 338 \pi^2 - 507 \psi^\prime(\third) + 864 \right] C_A^2 ~+~ 
\left[ 312 \psi^\prime(\third) - 208 \pi^2 - 108 \right] C_A T_F \Nf
\right. \nonumber \\
&& \left. ~~~+ \left[ 312 \psi^\prime(\third) - 208 \pi^2 - 2808 \right] 
C_A C_F \right. \nonumber \\
&& \left. ~~~+ \left[ 128 \pi^2 - 192 \psi^\prime(\third) + 3024 \right] 
C_F T_F \Nf \right] \frac{a^2}{324} \nonumber \\ 
&& +~ \left[ \left[ 
548808 \pi^2 \psi^\prime(\third) 
- 411606 (\psi^\prime(\third))^2
+ 4914 \psi^\prime(\third)
+ 27729 \psi^{\prime\prime\prime}(\third) 
\right. \right. \nonumber \\
&& \left. \left. ~~~~~
+~ 8643024 s_2(\pisix)
- 17286048 s_2(\pitwo)
- 14405040 s_3(\pisix)
+ 11524032 s_3(\pitwo)
\right. \right. \nonumber \\
&& \left. \left. ~~~~~
-~ 256880 \pi^4 
- 3276 \pi^2 
- 707616 \Sigma 
+ 1016226 \zeta(3) 
- 2542266
\right. \right. \nonumber \\
&& \left. \left. ~~~~~
+~ 60021 \frac{\ln^2(3) \pi}{\sqrt{3}} 
- 720252 \frac{\ln(3) \pi}{\sqrt{3}}
- 64467 \frac{\pi^3}{\sqrt{3}}
\right] C_A^3 \right. \nonumber \\
&& ~~~~~+ \left. \left[ 
745056 (\psi^\prime(\third))^2
- 993408 \pi^2 \psi^\prime(\third) 
- 8587296 \psi^\prime(\third)
+ 16848 \psi^{\prime\prime\prime}(\third) 
\right. \right. \nonumber \\
&& \left. \left. ~~~~~~~~~
-~ 25474176 s_2(\pisix)
+ 50948352 s_2(\pitwo)
+ 42456960 s_3(\pisix)
\right. \right. \nonumber \\
&& \left. \left. ~~~~~~~~~
-~ 33965568 s_3(\pitwo)
+ 286208 \pi^4 
+ 5724864 \pi^2 
+ 606528 \zeta(3) 
- 198288
\right. \right. \nonumber \\
&& \left. \left. ~~~~~~~~~
-~ 176904 \frac{\ln^2(3) \pi}{\sqrt{3}} 
+ 2122848 \frac{\ln(3) \pi}{\sqrt{3}}
+ 190008 \frac{\pi^3}{\sqrt{3}}
\right] C_A^2 C_F \right. \nonumber \\
&& ~~~~~+ \left. \left[ 
253296 (\psi^\prime(\third))^2
- 337728 \pi^2 \psi^\prime(\third) 
+ 1856304 \psi^\prime(\third)
- 28296 \psi^{\prime\prime\prime}(\third) 
\right. \right. \nonumber \\
&& \left. \left. ~~~~~~~~~
-~ 2892672 s_2(\pisix)
+ 5785344 s_2(\pitwo)
+ 4821120 s_3(\pisix)
- 3856896 s_3(\pitwo)
\right. \right. \nonumber \\
&& \left. \left. ~~~~~~~~~
+~ 188032 \pi^4 
- 1237536 \pi^2 
+ 435456 \Sigma 
+ 3059856 \zeta(3) 
+ 1166400
\right. \right. \nonumber \\
&& \left. \left. ~~~~~~~~~
-~ 20088 \frac{\ln^2(3) \pi}{\sqrt{3}} 
+ 241056 \frac{\ln(3) \pi}{\sqrt{3}}
+ 21576 \frac{\pi^3}{\sqrt{3}}
\right] C_A^2 T_F \Nf \right. \nonumber \\
&& ~~~~~+ \left. \left[ 
499200 \pi^2 \psi^\prime(\third) 
- 374400 (\psi^\prime(\third))^2
+ 5997888 \psi^\prime(\third)
- 101088 \psi^{\prime\prime\prime}(\third) 
\right. \right. \nonumber \\
&& \left. \left. ~~~~~~~~~
-~ 4852224 s_2(\pisix)
+ 9704448 s_2(\pitwo)
+ 8087040 s_3(\pisix)
- 6469632 s_3(\pitwo)
\right. \right. \nonumber \\
&& \left. \left. ~~~~~~~~~
+~ 103168 \pi^4 
- 3998592 \pi^2 
+ 404352 \Sigma 
+ 5660928 \zeta(3) 
- 6368544
\right. \right. \nonumber \\
&& \left. \left. ~~~~~~~~~
-~ 33696 \frac{\ln^2(3) \pi}{\sqrt{3}} 
+ 404352 \frac{\ln(3) \pi}{\sqrt{3}}
+ 36192 \frac{\pi^3}{\sqrt{3}}
\right] C_A C_F^2 \right. \nonumber \\
&& ~~~~~+ \left. \left[ 
611328 \pi^2 \psi^\prime(\third) 
- 458496 (\psi^\prime(\third))^2
+ 5660928 \psi^\prime(\third)
+ 10368 \psi^{\prime\prime\prime}(\third) 
\right. \right. \nonumber \\
&& \left. \left. ~~~~~~~~~
+~ 15676416 s_2(\pisix)
- 31352832 s_2(\pitwo)
- 26127360 s_3(\pisix)
\right. \right. \nonumber \\
&& \left. \left. ~~~~~~~~~
+~ 20901888 s_3(\pitwo)
- 176128 \pi^4 
- 3773952 \pi^2 
\right. \right. \nonumber \\
&& \left. \left. ~~~~~~~~~
-~ 8584704 \zeta(3) 
+ 7309440
+ 108864 \frac{\ln^2(3) \pi}{\sqrt{3}} 
\right. \right. \nonumber \\
&& \left. \left. ~~~~~~~~~
-~ 1306368 \frac{\ln(3) \pi}{\sqrt{3}}
- 116928 \frac{\pi^3}{\sqrt{3}}
\right] C_A C_F T_F \Nf \right. \nonumber \\
&& ~~~~~+ \left. \left[ 
6912 \psi^{\prime\prime\prime}(\third) 
- 981504 \psi^\prime(\third)
- 1492992 s_2(\pisix)
+ 2985984 s_2(\pitwo)
\right. \right. \nonumber \\
&& \left. \left. ~~~~~~~~~
+~ 2488320 s_3(\pisix)
- 1990656 s_3(\pitwo)
- 18432 \pi^4 
+ 654336 \pi^2 
\right. \right. \nonumber \\
&& \left. \left. ~~~~~~~~~
-~ 1244160 \zeta(3) 
+ 155520
+ 10368 \frac{\ln^2(3) \pi}{\sqrt{3}} 
\right. \right. \nonumber \\
&& \left. \left. ~~~~~~~~~
+~ 124416 \frac{\ln(3) \pi}{\sqrt{3}}
+ 11136 \frac{\pi^3}{\sqrt{3}}
\right] C_A T_F^2 \Nf^2 \right. \nonumber \\
&& ~~~~~+ \left. \left[ 
230400 \pi^2 \psi^\prime(\third) 
- 307200 (\psi^\prime(\third))^2
- 4188672 \psi^\prime(\third)
+ 62208 \psi^{\prime\prime\prime}(\third) 
\right. \right. \nonumber \\
&& \left. \left. ~~~~~~~~~
+~ 2985984 s_2(\pisix)
- 597198 s_2(\pitwo)
- 4976640 s_3(\pisix)
+ 3981312 s_3(\pitwo)
\right. \right. \nonumber \\
&& \left. \left. ~~~~~~~~~
-~ 63488 \pi^4 
+ 2792448 \pi^2 
- 248832 \Sigma 
- 3483648 \zeta(3) 
+ 8118144
\right. \right. \nonumber \\
&& \left. \left. ~~~~~~~~~
+~ 20736 \frac{\ln^2(3) \pi}{\sqrt{3}} 
- 248832 \frac{\ln(3) \pi}{\sqrt{3}}
- 22272 \frac{\pi^3}{\sqrt{3}}
\right] C_F^2 T_F \Nf \right. \nonumber \\
&& ~~~~~+ \left. \left[ 
165888 \psi^\prime(\third)
- 110592 \pi^2 
+ 2985984 \zeta(3) \right. \right. \nonumber \\
&& \left. \left. ~~~~~~~~~  
-~ 3608064 \right] C_F T_F^2 \Nf^2 \right] \frac{a^3}{139968} ~+~ O(a^4)
\end{eqnarray}
\begin{eqnarray}
\gamma_c^{\mbox{$\MOMqs$}}(a,0) &=& -~ \frac{3}{4} C_A a \nonumber \\
&& +~ \left[ \left[ 26 \pi^2 - 39 \psi^\prime(\third) + 90 \right] C_A ~+~ 
36 T_F \Nf \right. \nonumber \\
&& \left. ~~~~+ \left[ 24 \psi^\prime(\third) - 16 \pi^2 - 216 \right] 
C_F \right] \frac{C_A a^2}{72} \nonumber \\ 
&& +~ \left[ \left[ 
42216 \pi^2 \psi^\prime(\third) 
- 31662 (\psi^\prime(\third))^2
+ 15066 \psi^\prime(\third)
+ 2133 \psi^{\prime\prime\prime}(\third) 
\right. \right. \nonumber \\
&& \left. \left. ~~~~~
+~ 664848 s_2(\pisix)
- 1329696 s_2(\pitwo)
- 1108080 s_3(\pisix)
+ 886464 s_3(\pitwo)
\right. \right. \nonumber \\
&& \left. \left. ~~~~~
-~ 19760 \pi^4 
-~ 10044 \pi^2 
- 54432 \Sigma 
+ 40338 \zeta(3) 
- 267786
\right. \right. \nonumber \\
&& \left. \left. ~~~~~
+~ 4617 \frac{\ln^2(3) \pi}{\sqrt{3}} 
- 55404 \frac{\ln(3) \pi}{\sqrt{3}}
- 4959 \frac{\pi^3}{\sqrt{3}}
\right] C_A^2 \right. \nonumber \\
&& ~~~~~+ \left. \left[ 
57312 (\psi^\prime(\third))^2
- 76416 \pi^2 \psi^\prime(\third) 
- 669600 \psi^\prime(\third)
+ 1296 \psi^{\prime\prime\prime}(\third) 
\right. \right. \nonumber \\
&& \left. \left. ~~~~~~~~~
-~ 1959552 s_2(\pisix)
+ 3919104 s_2(\pitwo)
+ 3265920 s_3(\pisix)
\right. \right. \nonumber \\
&& \left. \left. ~~~~~~~~~
-~ 2612736 s_3(\pitwo)
+ 22016 \pi^4 
+ 446400 \pi^2 
+ 46656 \zeta(3) 
+ 66096
\right. \right. \nonumber \\
&& \left. \left. ~~~~~~~~~
-~ 13608 \frac{\ln^2(3) \pi}{\sqrt{3}} 
+ 163296 \frac{\ln(3) \pi}{\sqrt{3}}
+ 14616 \frac{\pi^3}{\sqrt{3}}
\right] C_A C_F \right. \nonumber \\
&& ~~~~~+ \left. \left[ 
38400 (\psi^\prime(\third))^2
- 28800 \pi^2 \psi^\prime(\third) 
+ 461376 \psi^\prime(\third)
- 7776 \psi^{\prime\prime\prime}(\third) 
\right. \right. \nonumber \\
&& \left. \left. ~~~~~~~~~
-~ 373248 s_2(\pisix)
+ 746496 s_2(\pitwo)
+ 622080 s_3(\pisix)
- 497664 s_3(\pitwo)
\right. \right. \nonumber \\
&& \left. \left. ~~~~~~~~~
+~ 7936 \pi^4 
-~ 307584 \pi^2 
+ 31104 \Sigma 
+ 435456 \zeta(3) 
- 489888
\right. \right. \nonumber \\
&& \left. \left. ~~~~~~~~~
-~ 2592 \frac{\ln^2(3) \pi}{\sqrt{3}} 
+ 31104 \frac{\ln(3) \pi}{\sqrt{3}}
+ 2784 \frac{\pi^3}{\sqrt{3}}
\right] C_F^2 ~-~ 
103680 T_F^2 \Nf^2 \right. \nonumber \\
&& ~~~~~+ \left. \left[ 
145152 \psi^\prime(\third)
- 864 \psi^{\prime\prime\prime}(\third) 
+ 186624 s_2(\pisix)
- 373248 s_2(\pitwo)
\right. \right. \nonumber \\
&& \left. \left. ~~~~~~~~~
-~ 311040 s_3(\pisix)
+ 248832 s_3(\pitwo)
+ 2304 \pi^4 
- 96768 \pi^2 
- 15552 \zeta(3) 
\right. \right. \nonumber \\
&& \left. \left. ~~~~~~~~~
+~ 273456
+ 1296 \frac{\ln^2(3) \pi}{\sqrt{3}} 
- 15552 \frac{\ln(3) \pi}{\sqrt{3}}
- 1392 \frac{\pi^3}{\sqrt{3}}
\right] C_A T_F \Nf \right. \nonumber \\
&& \left. ~~~~~+~ 
\left[ 
23040 \pi^2 
- 34560 \psi^\prime(\third)
+ 264384
\right] C_F T_F \Nf \right] \frac{C_A a^3}{31104} ~+~ O(a^4) 
\end{eqnarray}
and 
\begin{eqnarray}
\gamma_\psi^{\mbox{$\MOMqs$}}(a,0) &=& \left[ 25 C_A - 6 C_F 
- 8 T_F \Nf \right] \frac{C_F a^2}{4} \nonumber \\
&& +~ \left[ \left[ 3900 \psi^\prime(\third) - 2600 \pi^2
- 13230 \zeta(3) + 10287 \right] C_A^2 \right. \nonumber \\
&& \left. ~~~~+ \left[ 2224 \pi^2 - 3336 \psi^\prime(\third) 
+ 5184 \zeta(3) + 16092 \right] C_A C_F \right. \nonumber \\
&& \left. ~~~~+ \left[ 832 \pi^2 - 1248 \psi^\prime(\third) 
+ 3456 \zeta(3) - 8784 \right] C_A T_F \Nf \right. \nonumber \\
&& \left. ~~~~+ \left[ 768 \psi^\prime(\third) - 512 \pi^2 
- 7776 \right] C_F T_F \Nf ~+~ 1152 T_F^2 \Nf^2 \right. \nonumber \\
&& \left. ~~~~+ \left[ 576 \psi^\prime(\third) - 384 \pi^2 - 4536 \right] C_F^2
\right] \frac{C_F a^2}{432} ~+~ O(a^4) ~.
\end{eqnarray}
The corresponding arbitrary linear covariant $SU(3)$ expressions are 
\begin{eqnarray}
\beta^{\mbox{$\MOMqs$}}(a,\alpha) &=& -~ [ 11.0000000 - 0.6666667 \Nf ] a^2
\nonumber \\
&& -~ \left[ 102.0000000 + 15.2372141 \alpha - 1.3839787 \alpha^2
- 0.4920696 \alpha^3 \right. \nonumber \\
&& \left. ~~~~- \left[ 12.6666667 + 1.5627912 \alpha + 0.2186976 \alpha^2
\right] \Nf \right] a^3 \nonumber \\
&& -~ \left[ 1843.6527285 + 422.0731852 \alpha + 123.3734958 \alpha^2
- 19.5130255 \alpha^3 \right. \nonumber \\
&& \left. ~~~~- 3.5055190 \alpha^4 - 0.0961314 \alpha^5 \right. \nonumber \\
&& \left. ~~~~- \left[ 588.6548455 + 60.5454812 \alpha + 16.3955703 \alpha^2
+ 0.9282359 \alpha^3 \right. \right. \nonumber \\
&& \left. \left. ~~~~~~~~- 0.0000059 \alpha^4 \right] \Nf ~+~ 
22.5878118 \Nf^2 \right] a^4 ~+~ O(a^5) \nonumber \\
\gamma_A^{\mbox{$\MOMqs$}}(a,\alpha) &=& [ 0.6666667 \Nf - 6.5000000
+ 1.5000000 \alpha ] a \nonumber \\
&& -~ \left[ 46.6391320 + 22.5608759 \alpha - 6.2001294 \alpha^2
+ 0.8789652 \alpha^3 \right. \nonumber \\
&& \left. ~~~~- \left[ 9.4117058 + 1.5627912 \alpha - 0.3906512 \alpha^2
 \right] \Nf \right] a^2 \nonumber \\
&& -~ \left[ 2027.7437143 + 333.3082218 \alpha + 184.2382915 \alpha^2
- 24.3519716 \alpha^3 \right. \nonumber \\
&& \left. ~~~~+ 12.6718858 \alpha^4 + 1.9200786 \alpha^5 \right. \nonumber \\
&& \left. ~~~~- \left[ 415.6990152 + 49.4053081 \alpha + 9.7003844 \alpha^2
- 3.7908552 \alpha^3 \right. \right. \nonumber \\
&& \left. \left. ~~~~~~~~- 0.4783683 \alpha^4 \right] \Nf \right. \nonumber \\
&& \left. ~~~~+ \left[ 11.1788079 - 1.3021707 \alpha \right] \Nf^2 \right] 
a^3 ~+~ O(a^4) \nonumber \\
\gamma_c^{\mbox{$\MOMqs$}}(a,\alpha) &=& [ 0.7500000 \alpha - 2.2500000 ] a 
\nonumber \\
&& -~ \left[ 13.2020072 + 12.3112512 \alpha - 2.5140879 \alpha^2
- 0.6855174 \alpha^3 \right. \nonumber \\
&& \left. ~~~~-~ 0.7500000 \Nf \right] a^2 \nonumber \\
&& -~ \left[ 740.1341645 + 1.8665778 \alpha + 100.6450352 \alpha^2
- 3.4355918 \alpha^3 \right. \nonumber \\
&& \left. ~~~~- 8.7678000 \alpha^4 - 1.0965129 \alpha^5 \right. \nonumber \\
&& \left. ~~~~-~ \left[ 75.5032720 + 4.1186469 \alpha + 1.7109768 \alpha^2 
\right] \Nf + 2.5000000 \Nf^2 \right] a^3 \nonumber \\
&& +~ O(a^4) \nonumber \\
\gamma_\psi^{\mbox{$\MOMqs$}}(a,\alpha) &=& 1.3333333 \alpha a \nonumber \\
&& +~ \left[ 22.3333333 + 2.4900784 \alpha + 8.1255824 \alpha^2
+ 1.2186976 \alpha^3 \right. \nonumber \\
&& \left. ~~~~- 1.3333333 \Nf \right] a^2 \nonumber \\
&& +~ \left[ 341.8989103 + 182.9132891 \alpha + 43.9801061 \alpha^2
+ 74.9283461 \alpha^3 \right. \nonumber \\
&& \left. ~~~~+ 21.4352687 \alpha^4 + 1.9493562 \alpha^5 \right. \nonumber \\
&& \left. ~~~~- \left[ 52.1916907 - 3.1076278 \alpha - 2.1669462 \alpha^2
\right] \Nf ~+~ 0.8888889 \Nf^2 \right] a^3 \nonumber \\
&& +~ O(a^4) ~.  
\end{eqnarray}

Unlike the previous two sections we close this section with a few remarks on
other renormalization group functions in QCD. Recently, there has been interest
in the RI${}^\prime$/SMOM renormalization scheme which was developed in 
\cite{49}. Briefly the scheme was introduced to renormalize flavour non-singlet 
quark currents in a quark $2$-point function at the symmetric subtraction 
point. The motivation for such a scheme is that it circumvents potential 
infrared singularities of the measurement of the same Green's function on the 
lattice in the chiral limit for exceptional momentum configurations, \cite{20}.
More specifically the scalar, vector and tensor currents have been considered 
at one and two loops in this scheme, \cite{49,43,50}, and the associated 
conversion functions for each operator relative to the $\MSbar$ scheme have
been determined. Thus the RI${}^\prime$/SMOM three loop operator anomalous 
dimensions have been deduced in an arbitrary covariant gauge. More recently, 
the full set of amplitudes has been provided for each of these currents at two 
loops in \cite{51}. Equally the same information has been provided for the 
$n$~$=$~$2$ and $3$ moments of the flavour non-singlet Wilson operator central 
to deep inelastic scattering, \cite{52,53}. As an extension of the 
RI${}^\prime$/SMOM analysis we have determined the three loop anomalous 
dimensions of the scalar and tensor quark currents as well as the 
$n$~$=$~$2$ moment of the Wilson operator in the MOMq scheme. Numerically the 
$SU(3)$ values are  
\begin{eqnarray}
\gamma_S^{\mbox{$\MOMqs$}}(a,\alpha) &=& -~ 4.0000000 a - \left[ 7.5709424 
+ 8.3450254 \alpha + 0.3121855 \alpha^2 + 1.7918763 \Nf \right] a^2 
\nonumber \\
&& +~ \left[ 324.9490278 + 84.8639648 \alpha - 37.0694357 \alpha^2
- 1.1713670 \alpha^3 \right. \nonumber \\
&& \left. ~~~~+ 0.2647488 \alpha^4 - \left[ 16.2843867 + 18.7759399 \alpha 
+ 0.5878184 \alpha^2 \right] \Nf \right. \nonumber \\
&& \left. ~~~~- 2.6666667 \Nf^2 \right] a^3 ~+~ O(a^4) 
\label{anomscal}
\end{eqnarray}
\begin{eqnarray}
\gamma_T^{\mbox{$\MOMqs$}}(a,\alpha) &=& 1.3333333 a + \left[ 20.3014252 
+ 8.1573041 \alpha + 1.8959382 \alpha^2 - 0.5878931 \Nf \right] a^2 
\nonumber \\
&& +~ \left[ 276.1622997 + 149.4337760 \alpha + 83.3978821 \alpha^2
+ 32.6152122 \alpha^3 \right. \nonumber \\
&& \left. ~~~~+ 3.1873834 \alpha^4 - \left[ 61.3516342 + 4.4008934 \alpha 
+ 0.1928562 \alpha^2 \right] \Nf \right. \nonumber \\
&& \left. ~~~~+ 2.0740741 \Nf^2 \right] a^3 ~+~ O(a^4) 
\label{anomtens}
\end{eqnarray}
and
\begin{eqnarray}
\gamma_{W_2,11}^{\mbox{$\MOMqs$}}(a,\alpha) &=& 3.5555556 a + \left[ 
45.2685593 + 15.7105153 \alpha + 3.0417366 \alpha^2 - 2.6264345 \Nf \right] a^2 
\nonumber \\
&& +~ \left[ 818.7944256 + 154.0549414 \alpha + 59.1199636 \alpha^2
+ 6.2844353 \alpha^3 \right. \nonumber \\
&& \left. ~~~~+ 0.3234930 \alpha^4 - \left[ 140.9726462 + 5.3687957 \alpha 
+ 0.8615923 \alpha^2 \right] \Nf \right. \nonumber \\
&& \left. ~~~~+ 4.7392379 \Nf^2 \right] a^3 ~+~ O(a^4) \nonumber \\ 
\gamma_{W_2,12}^{\mbox{$\MOMqs$}}(a,\alpha) &=& -~ 1.7777778 a - \left[ 
16.5859286 + 7.5010250 \alpha + 0.9582634 \alpha^2 - 0.7715739 \Nf \right] a^2 
\nonumber \\
&& -~ \left[ 298.1271156 + 48.6705419 \alpha + 27.5758422 \alpha^2
+ 3.1422177 \alpha^3 \right. \nonumber \\
&& \left. ~~~~+ 0.1617465 \alpha^4 - \left[ 54.9281680 + 0.1449714 \alpha 
+ 0.2531120 \alpha^2 \right] \Nf \right. \nonumber \\
&& \left. ~~~~+ 2.1752211 \Nf^2 \right] a^3 ~+~ O(a^4) \nonumber \\ 
\gamma_{W_2,22}^{\mbox{$\MOMqs$}}(a,\alpha) &=& O(a^4)
\label{anomw2}
\end{eqnarray}
where $S$ and $T$ respectively denote scalar and tensor currents. The Wilson
operator second moment three loop anomalous dimension corresponds to the $11$
entry of the dimension two upper triangular mixing matrix, \cite{52}. The $22$ 
entry corresponds to the total derivative operator for this operator moment but
its anomalous dimension is equivalent to that of the vector current, \cite{52}.
As the vector current is conserved and hence a physical operator its anomalous
dimension is zero to all orders in perturbation theory consistent with the 
Slavnov-Taylor identity. Whilst we have included a subset of the operators
considered in recent RI${}^\prime$/SMOM analyses for comparison we have not 
recorded the complete set for the MOMq or either of the other MOM schemes
discussed here. The reason for this is simple. In deriving (\ref{anomscal}),
(\ref{anomtens}) and (\ref{anomw2}) we have arrived at the results by 
constructing the respective conversion functions and then using the analogous
relation to (\ref{anomdef}) for operator renormalization. It transpires, like
our observation for the wave function and gauge parameter conversion functions,
that the respective conversion functions are formally the same as those 
already given in \cite{49,43,50} for the quark currents and that for $W_2$ 
given in \cite{52}. The only difference here in producing (\ref{anomscal}), 
(\ref{anomtens}) and (\ref{anomw2}) is that the coupling constant mapping 
between the schemes is not trivial as it is in the RI${}^\prime$/SMOM case. 
Therefore, if one is interested in the structure of the three loop anomalous 
dimensions in the MOMggg and MOMh schemes it is merely a simple exercise to 
construct them from the known conversion functions. We should add that in 
deriving the two loop parts of (\ref{anomscal}), (\ref{anomtens}) and 
(\ref{anomw2}) we have not only used the renormalization group construction but
also carried out the explicit two loop renormalization of each of the three 
operators in the MOMq scheme. Therefore, the two loop terms of each have been 
derived directly and also indirectly from the respective one loop conversion 
functions which we have constructed explicitly.

\sect{Discussion.} 

We close with some remarks and give an overall perspective on our computations.
First, we have computed the two loop structure of each of the trivalent QCD 
vertices with a linear covariant gauge fixing at the symmetric subtraction 
point for the $\MSbar$ scheme as well as the three MOM schemes. Although we 
have only given the MOMggg, MOMh and MOMq versions for each of the respective 
triple gluon, ghost-gluon and quark-gluon vertices, the values of say the 
triple gluon vertex in the MOMq scheme, if of interest, can be readily deduced 
from the results given here via the Slavnov-Taylor identities of QCD, 
\cite{21}. In addition we have deduced the basic three loop renormalization 
group functions including the $\beta$-functions in each of the three MOM 
schemes in an arbitrary linear covariant gauge. As there has been numerical 
estimates for some of these quantities in the Landau gauge in \cite{22} we need
to comment on how our results compare in addition to earlier remarks. First, we 
consider the $SU(3)$ Landau gauge two loop mapping of the coupling constant 
from the MOMi schemes to $\MSbar$. Specifically, the polynomials in $\Nf$ of 
the perturbative map were given in \cite{22} and the coefficients were denoted 
by $d_{lj}^{\MOMis}$, where $l$ is the loop order and $j$ is the power of $\Nf$
in the polynomial. As $d_{10}^{\MOMis}$, $d_{11}^{\MOMis}$ and 
$d_{22}^{\MOMis}$ can all be computed analytically without approximation and 
which we agree with for MOMggg and MOMq, we focus on the remaining two, 
$d_{20}^{\MOMis}$ and $d_{21}^{\MOMis}$, which could only be evaluated 
numerically in \cite{22}. In Table $3$ we give the results of \cite{22} in its 
notation, in order to preserve the error estimates, together with the numerical
evaluation of our analytic results for $SU(3)$ in the same convention. Aside 
from the MOMh scheme the agreement is remarkably close. However, that for 
$d_{21}^{\MOMhs}$ differs by about $5$\% whilst that for $d_{20}^{\MOMhs}$ is 
significantly different. We believe that this is a transcription inconsistency 
in the presentation of the results in \cite{22} akin to that noted earlier here
and in \cite{28} for the amplitudes of this vertex. This is also because we do 
not tally with the one loop coefficients $d_{10}^{\MOMhs}$ and 
$d_{11}^{\MOMhs}$ which are known exactly, \cite{21}. We have checked that we 
obtain the {\em same} one loop MOMh ghost-gluon vertex counterterm given in 
\cite{21} {\em exactly} for all $\alpha$. Equally we have obtained the same 
vertex counterterm as \cite{21} for the quark-gluon vertex as a check on our 
overall procedures which were the same for each vertex. For instance, the one 
loop master integral denoted by $I$ in \cite{21} and \cite{22} arises in 
$d_{10}^{\MOMhs}$. It derives directly from the vertex counterterm only and the
coefficient of $I$ is therefore in a one-to-one mapping to the coefficient of 
$I$ in $d_{10}^{\MOMhs}$. Comparing the counterterms of our results and 
\cite{22}, it is clear that the coefficients differ but we find our expressions
are consistent with \cite{21}. Therefore, the one loop agreement of our MOMh 
results with \cite{21} suggests that we have a more credible MOMh coupling 
constant mapping as we have followed the same symbolic algebraic algorithm as 
that for MOMggg and MOMq.

{\begin{table}[ht]
\begin{center}
\begin{tabular}{|c||c|c||c|c||c|c|}
\hline
& $d_{20}^{\MOMgggs}$ & $d_{21}^{\MOMgggs}$ & $d_{20}^{\MOMhs}$ & 
$d_{21}^{\MOMhs}$ & $d_{20}^{\MOMqs}$ & $d_{21}^{\MOMqs}$ \\ 
\hline
Ref \cite{22} & $59.8(8)$ & $-$ $12.6(2)$ & $35.88(2)$ & $-$ $5.0707(6)$ 
& $29.53(1)$ & $-$ $5.1961(4)$ \\
This paper & $60.02892$ & $-$ $12.63031$ & $40.12125$ & $-$ $5.34725$ 
& $29.50994$ & $-$ $5.19445$ \\
\hline
\end{tabular}
\end{center}
\begin{center}
{Table $3$. Comparison of two of the two loop $SU(3)$ coupling constant 
relation coefficients for each scheme in the notation and convention of 
\cite{22}.}
\end{center}
\end{table}}

Concerning the final Landau gauge $SU(3)$ $\beta$-functions of \cite{22}, we 
have already briefly noted the accuracy of our exact expressions in \cite{24}.
Summarizing, for the coefficient of the $\Nf$ independent three loop term the 
agreement was around $2$\% for the MOMggg and MOMh $\beta$-functions and less 
than $0.2$\% for MOMq. In Table $4$ we have presented the coefficients of the 
polynomial in $\Nf$ of the three loop term of the $\beta$-function in each
scheme. These are denoted by $\beta_{lj}$ where again $l$ is the loop order and
$j$ is the power of $\Nf$ in the polynomial. Only the MOMggg scheme has a
cubic term in $\Nf$ and it is known analytically. So the precise numerical
agreement merely reflects this. The linear term in $\Nf$ in the MOMggg scheme 
is virtually zero because of an accidental cancellation for the colour group 
$SU(3)$ which was noted in \cite{22}. This is the reason why a large error was 
recorded for this coefficient though we obtain the same sign for our 
coefficient. Overall for MOMggg and MOMq all the coefficients are very close to
the central values given in \cite{22}. That for $\beta_{30}^{\MOMgggs}$ had a
relatively large error, possibly as a result of the purely gluonic graphs which
might have a more slow convergence in the asymmetric expansion parameter
approach used in \cite{22}. However, it is still reassuring that our result 
emerges so close to the central value in this particular case which roughly has
a $2$\% error. Therefore, in this context the slightly smaller error on 
$\beta_{30}^{\MOMhs}$ suggests that ultimately the MOMh $\beta$-function of 
\cite{22} has been correctly obtained. Although the agreement of the other 
$\Nf$ polynomial coefficients is reasonably decent there does not appear to be 
as close an overlap compared to say those for MOMq. This is probably, because, 
as we noted earlier, the amplitude for the triple gluon and quark-gluon 
vertices for the channel containing the poles in $\epsilon$, or equivalently 
the channel corresponding to the vertex Feynman rule, are in very good 
agreement with the Landau gauge estimates given in \cite{22}. Therefore, 
overall we are confident that our results are correct.

{\begin{table}[ht]
\begin{center}
\begin{tabular}{|c|c||c|c|c|c|}
\hline
Scheme & Article & $\beta_{30}$ & $\beta_{31}$ & $\beta_{32}$ & $\beta_{33}$ \\ 
\hline
MOMggg & Ref \cite{22} & $24(2)$ & $0.04(63)$ & $-$ $1.05(3)$ & $0.0415330$ \\
& This paper & $24.5466309$ & $0.0088426$ & $-$ $1.0482740$ & $0.0415330$ \\
\hline
\hline
MOMh & Ref \cite{22} & $44.82(5)$ & $-$ $9.730(5)$ & $0.3276(1)$ & - \\
& This paper & $43.9608273$ & $-$ $9.6507368$ & $0.3359815$ & - \\
\hline
\hline
MOMq & Ref \cite{22} & $28.86(3)$ & $-$ $9.206(3)$ & $0.35322(7)$ & - \\
& This paper & $28.8070739$ & $-$ $9.1977320$ & $0.3529346$ & - \\
\hline
\hline
\end{tabular}
\end{center}
\begin{center}
{Table $4$. Comparison of coefficients of three loop Landau gauge 
$\beta$-function in the convention of \cite{22}.}
\end{center}
\end{table}}

One final point worth emphasising in this regard is to note that when the 
computations of \cite{22} were being carried out the technology available in 
terms of algorithms, such as \cite{40}, and computer power was not comparable 
to that of current levels. Therefore it is a remarkable achievement that
overall the results are in solid argreement. This is partly because the 
expansion technique of \cite{22} to evaluate a $3$-point Green's function at 
the symmetric point in terms of $2$-point functions to allow the application of 
{\sc Mincer}, requires a large amount of integration by parts especially as the
powers of the propagators increases in each term of the expansion. Despite 
this, however, it would seem that the method of \cite{22} could be extended in
principle to tackle the construction of the vertex functions to three loops. 
Whilst there are now three loop anomalous dimensions and $\beta$-functions 
available as intermediate checks on any numerical estimates, it would provide a
reasonable estimate for {\em four} loop renormalization group functions 
including the $\beta$-functions in physical MOM schemes. Such higher order 
computations may be necessary soon in order to not only assist with lattice 
matching of vertex functions but also for more accurate determination of the 
value of the strong coupling constant. 

\vspace{1cm}
\noindent
{\bf Acknowledgement.} The author thanks Prof. K.G. Chetyrkin for discussion on
\cite{22} and providing a copy of \cite{23}.

\appendix

\sect{Tensor basis and projectors.}

In this appendix we list in turn the basis of tensors used as well as the 
coefficients for the projection of each of the three vertices. Away from the
symmetric subtraction point the tensor basis is enlarged.

\subsection{Triple gluon vertex.}

For the triple gluon vertex we have the basis tensors 
\begin{eqnarray}
{\cal P}^{\mbox{\footnotesize{ggg}}}_{(1) \mu \nu \sigma }(p,q) &=& 
\eta_{\mu \nu} p_\sigma ~~,~~ 
{\cal P}^{\mbox{\footnotesize{ggg}}}_{(2) \mu \nu \sigma }(p,q) ~=~ 
\eta_{\nu \sigma} p_\mu ~~,~~ 
{\cal P}^{\mbox{\footnotesize{ggg}}}_{(3) \mu \nu \sigma }(p,q) ~=~ 
\eta_{\sigma \mu} p_\nu \nonumber \\
{\cal P}^{\mbox{\footnotesize{ggg}}}_{(4) \mu \nu \sigma }(p,q) &=& 
\eta_{\mu \nu} q_\sigma ~~,~~ 
{\cal P}^{\mbox{\footnotesize{ggg}}}_{(5) \mu \nu \sigma }(p,q) ~=~ 
\eta_{\nu \sigma} q_\mu ~~,~~ 
{\cal P}^{\mbox{\footnotesize{ggg}}}_{(6) \mu \nu \sigma }(p,q) ~=~ 
\eta_{\sigma \mu} q_\nu \nonumber \\
{\cal P}^{\mbox{\footnotesize{ggg}}}_{(7) \mu \nu \sigma }(p,q) &=& 
\frac{1}{\mu^2} p_\mu p_\nu p_\sigma ~~,~~ 
{\cal P}^{\mbox{\footnotesize{ggg}}}_{(8) \mu \nu \sigma }(p,q) ~=~ 
\frac{1}{\mu^2} p_\mu p_\nu q_\sigma ~~,~~ 
{\cal P}^{\mbox{\footnotesize{ggg}}}_{(9) \mu \nu \sigma }(p,q) ~=~ 
\frac{1}{\mu^2} p_\mu q_\nu p_\sigma \nonumber \\ 
{\cal P}^{\mbox{\footnotesize{ggg}}}_{(10) \mu \nu \sigma }(p,q) &=& 
\frac{1}{\mu^2} q_\mu p_\nu p_\sigma ~~,~~ 
{\cal P}^{\mbox{\footnotesize{ggg}}}_{(11) \mu \nu \sigma }(p,q) ~=~ 
\frac{1}{\mu^2} p_\mu q_\nu q_\sigma ~~,~~ 
{\cal P}^{\mbox{\footnotesize{ggg}}}_{(12) \mu \nu \sigma }(p,q) ~=~ 
\frac{1}{\mu^2} q_\mu p_\nu q_\sigma \nonumber \\ 
{\cal P}^{\mbox{\footnotesize{ggg}}}_{(13) \mu \nu \sigma }(p,q) &=& 
\frac{1}{\mu^2} q_\mu q_\nu p_\sigma ~~,~~ 
{\cal P}^{\mbox{\footnotesize{ggg}}}_{(14) \mu \nu \sigma }(p,q) ~=~ 
\frac{1}{\mu^2} q_\mu q_\nu q_\sigma ~. 
\end{eqnarray}
The first six tensors represent the structures which appear in the triple gluon
vertex Feynman rule. As there are fourteen tensors for this vertex then in
order to simplify the presentation we have partitioned the projection matrix 
into a $3$~$\times$~$3$ block matrix with partitions of sizes $6$, $4$ and $4$. 
Defining 
\begin{eqnarray}
{\cal M}^{\mbox{\footnotesize{ggg}}} &=& -~ \frac{1}{27(d-2)} \left(
\begin{array}{ccc}
{\cal M}^{\mbox{\footnotesize{ggg}}}_{11} & 
{\cal M}^{\mbox{\footnotesize{ggg}}}_{12} & 
{\cal M}^{\mbox{\footnotesize{ggg}}}_{13} \\
{\cal M}^{\mbox{\footnotesize{ggg}}}_{21} & 
{\cal M}^{\mbox{\footnotesize{ggg}}}_{22} & 
{\cal M}^{\mbox{\footnotesize{ggg}}}_{23} \\
{\cal M}^{\mbox{\footnotesize{ggg}}}_{31} & 
{\cal M}^{\mbox{\footnotesize{ggg}}}_{32} & 
{\cal M}^{\mbox{\footnotesize{ggg}}}_{33} \\
\end{array}
\right) \nonumber
\end{eqnarray}
then in $d$-dimensions the sub-matrices for the projection are 
\begin{eqnarray}
{\cal M}^{\mbox{\footnotesize{ggg}}}_{11} &=& \left(
\begin{array}{cccccc}
36 & 0 & 0 & 18 & 0 & 0 \\
0 & 36 & 0 & 0 & 18 & 0 \\
0 & 0 & 36 & 0 & 0 & 18 \\
18 & 0 & 0 & 36 & 0 & 0 \\
0 & 18 & 0 & 0 & 36 & 0 \\
0 & 0 & 18 & 0 & 0 & 36 \\
\end{array}
\right) ~,~
{\cal M}^{\mbox{\footnotesize{ggg}}}_{12} ~=~ \left(
\begin{array}{cccc}
48 & 24 & 24 & 24 \\
48 & 24 & 24 & 24 \\
48 & 24 & 24 & 24 \\
24 & 48 & 12 & 12 \\
24 & 12 & 12 & 48 \\
24 & 12 & 48 & 12 \\
\end{array}
\right) \nonumber \\
{\cal M}^{\mbox{\footnotesize{ggg}}}_{13} &=& \left(
\begin{array}{cccc}
12 & 12 & 48 & 24 \\
48 & 12 & 12 & 24 \\
12 & 48 & 12 & 24 \\
24 & 24 & 24 & 48 \\
24 & 24 & 24 & 48 \\
24 & 24 & 24 & 48 \\
\end{array}
\right) ~,~
{\cal M}^{\mbox{\footnotesize{ggg}}}_{21} ~=~ \left(
\begin{array}{cccccc}
48 & 48 & 48 & 24 & 24 & 24 \\
24 & 24 & 24 & 48 & 12 & 12 \\
24 & 24 & 24 & 12 & 12 & 48 \\
24 & 24 & 24 & 12 & 48 & 12 \\
\end{array}
\right) \nonumber \\
{\cal M}^{\mbox{\footnotesize{ggg}}}_{22} &=& \left(
\begin{array}{cccc}
64 (d+1) & 32 (d+1) & 32 (d+1) & 32 (d+1) \\ 
32 (d+1) & 32 (2d-1) & 16 (d+1) & 16 (d+1) \\ 
32 (d+1) & 16 (d+1) & 32 (2d-1) & 16 (d+1) \\ 
32 (d+1) & 16 (d+1) & 16 (d+1) & 32 (2d-1) \\ 
\end{array}
\right) \nonumber \\
{\cal M}^{\mbox{\footnotesize{ggg}}}_{23} &=& \left(
\begin{array}{cccc}
16 (d+4) & 16 (d+4) & 16 (d+4) & 8 (d+10) \\ 
8 (4d+1) & 8 (4d+1) & 8 (d+4) & 16 (d+4) \\ 
8 (4d+1) & 8 (d+4) & 8 (4d+1) & 16 (d+4) \\ 
8 (d+4) & 8 (4d+1) & 8 (4d+1) & 16 (d+4) \\ 
\end{array}
\right) \nonumber \\
{\cal M}^{\mbox{\footnotesize{ggg}}}_{31} &=& \left(
\begin{array}{cccccc}
12 & 48 & 12 & 24 & 24 & 24 \\
12 & 12 & 48 & 24 & 24 & 24 \\
48 & 12 & 12 & 24 & 24 & 24 \\
24 & 24 & 24 & 48 & 48 & 48 \\
\end{array}
\right) \nonumber \\
{\cal M}^{\mbox{\footnotesize{ggg}}}_{32} &=& \left(
\begin{array}{cccc}
16 (d+4) & 8 (4d+1) & 8 (4d+1) & 8 (d+4) \\ 
16 (d+4) & 8 (4d+1) & 8 (d+4) & 8 (4d+1) \\ 
16 (d+4) & 8 (d+4) & 8 (4d+1) & 8 (4d+1) \\ 
8 (d+10) & 16 (d+4) & 16 (d+4) & 16 (d+4) \\ 
\end{array}
\right) \nonumber \\
{\cal M}^{\mbox{\footnotesize{ggg}}}_{33} &=& \left(
\begin{array}{cccc}
32 (2d-1) & 16 (d+1) & 16 (d+1) & 32 (d+1) \\ 
16 (d+1) & 32 (2d-1) & 16 (d+1) & 32 (d+1) \\ 
16 (d+1) & 16 (d+1) & 32 (2d-1) & 32 (d+1) \\ 
32 (d+1) & 32 (d+1) & 32 (d+1) & 64 (d+1) \\ 
\end{array}
\right) 
\end{eqnarray}
in $d$-dimensions.

\subsection{Ghost-gluon vertex.}

At the symmetric subtraction point there are two basis tensors for the
ghost-gluon vertex which are  
\begin{equation}
{\cal P}^{\mbox{\footnotesize{ccg}}}_{(1) \sigma }(p,q) ~=~ p_\sigma ~~~,~~~
{\cal P}^{\mbox{\footnotesize{ccg}}}_{(2) \sigma }(p,q) ~=~ q_\sigma ~.
\end{equation} 
Hence, in this case the projection matrix is relatively simple and is given by
\begin{equation}
{\cal M}^{\mbox{\footnotesize{ccg}}} ~=~ -~ \frac{1}{3} \left(
\begin{array}{cc}
4 & 2 \\
2 & 4 \\
\end{array}
\right) 
\end{equation}
in $d$-dimensions.

\subsection{Quark-gluon vertex.}
The quark-gluon vertex tensor basis involves six independent tensors. These 
are the same as those used for the vector current insertion in a quark 
$2$-point function in \cite{51} and are given by 
\begin{eqnarray}
{\cal P}^{\mbox{\footnotesize{qqg}}}_{(1) \sigma }(p,q) &=& 
\gamma_\sigma ~~~,~~~
{\cal P}^{\mbox{\footnotesize{qqg}}}_{(2) \sigma }(p,q) ~=~ 
\frac{{p}_\sigma \pslash}{\mu^2} ~~~,~~~
{\cal P}^{\mbox{\footnotesize{qqg}}}_{(3) \sigma }(p,q) ~=~ 
\frac{{p}_\sigma \qslash}{\mu^2} ~, \nonumber \\
{\cal P}^{\mbox{\footnotesize{qqg}}}_{(4) \sigma }(p,q) &=& 
\frac{{q}_\sigma \pslash}{\mu^2} ~~~,~~~
{\cal P}^{\mbox{\footnotesize{qqg}}}_{(5) \sigma }(p,q) ~=~ 
\frac{{q}_\sigma \qslash}{\mu^2} ~~~,~~~
{\cal P}^{\mbox{\footnotesize{qqg}}}_{(6) \sigma }(p,q) ~=~ 
\frac{1}{\mu^2} \Gamma_{(3) \, \sigma p q} 
\end{eqnarray}
where the generalized $d$-dimensional $\gamma$-matrices denoted by
$\Gamma^{\mu_1 \ldots \mu_n}_{(n)}$ were defined earlier. Equally the same
projection tensor in $d$-dimensions emerges as in \cite{51} which is 
\begin{equation}
{\cal M}^{\mbox{\footnotesize{qqg}}} ~=~ \frac{1}{36(d-2)} \left(
\begin{array}{cccccc}
9 & 12 & 6 & 6 & 12 & 0 \\
12 & 16 (d - 1) &  8 (d - 1) &  8 (d - 1) & 4 (d + 2) & 0 \\
6 & 8 (d - 1) & 4 (4 d - 7) &  4 (d - 1) & 8 (d - 1) & 0 \\
6 & 8 (d - 1) &  4 (d - 1) & 4 (4 d - 7) & 8 (d - 1) & 0 \\
12 & 4 (d + 2) &  8 (d - 1) &  8 (d - 1) & 16 (d - 1) & 0 \\
0 & 0 & 0 & 0 & 0 & - 12 \\
\end{array}
\right) ~.
\end{equation}
The partition of the projection matrix into the $\Gamma_{(1)}^\mu$ and 
$\Gamma_{(3)}^{\mu\nu\sigma}$ sectors is apparent.

\end{document}